\documentclass[fleqn,usenatbib]{mnras}

\usepackage{newtxtext,newtxmath}

\usepackage[T1]{fontenc}

\DeclareRobustCommand{\VAN}[3]{#2}
\let\VANthebibliography\thebibliography
\def\thebibliography{\DeclareRobustCommand{\VAN}[3]{##3}\VANthebibliography}

\usepackage{graphicx}	
\usepackage{adjustbox}  
\usepackage{amsmath}	
\usepackage{wasysym}    
\usepackage{bm}         
\usepackage{xspace}
\usepackage{xcolor}
\makeatletter 
  \patchcmd{\NAT@citex}
    {\@citea\NAT@hyper@{%
      \NAT@nmfmt{\NAT@nm}%
      \hyper@natlinkbreak{\NAT@aysep\NAT@spacechar}{\@citeb\@extra@b@citeb}%
      \NAT@date}}
    {\@citea\NAT@nmfmt{\NAT@nm}%
    \NAT@aysep\NAT@spacechar\NAT@hyper@{\NAT@date}}{}{}

  \patchcmd{\NAT@citex}
    {\@citea\NAT@hyper@{%
      \NAT@nmfmt{\NAT@nm}%
      \hyper@natlinkbreak{\NAT@spacechar\NAT@@open\if*#1*\else#1\NAT@spacechar\fi}%
        {\@citeb\@extra@b@citeb}%
      \NAT@date}}
    {\@citea\NAT@nmfmt{\NAT@nm}%
    \NAT@spacechar\NAT@@open\if*#1*\else#1\NAT@spacechar\fi\NAT@hyper@{\NAT@date}}
    {}{}
\makeatother

\newcommand\Msun{\mathrm{M}_{\astrosun}} 
\newcommand\Zsun{\mathrm{Z}_{\astrosun}} 
\newcommand\HI{\ion{H}{I}\xspace} 
\newcommand\HII{\ion{H}{II}\xspace} 
\newcommand\OII{\ion{O}{II}\xspace} 
\newcommand\OIII{\ion{O}{III}\xspace} 

\title[The nature of diffuse ionised gas]{The nature of diffuse ionised gas in star-forming galaxies}

\author[W. McClymont et al.]{William McClymont,$^{1,2}$\thanks{E-mail: \href{mailto:wjm50@cam.ac.uk}{wjm50@cam.ac.uk} (WM)}
Sandro Tacchella,$^{1,2}$
Aaron Smith,$^{3}$
Rahul Kannan,$^{4}$
Roberto Maiolino,$^{1,2}$
\newauthor
Francesco Belfiore,$^{5}$
Lars Hernquist,$^{6}$
Hui Li$^{7}$
and Mark Vogelsberger$^{8}$
\\%
\\%
$^{1}$Kavli Institute for Cosmology, University of Cambridge, Madingley Road, Cambridge CB3 0HA, UK\\%
$^{2}$Cavendish Laboratory, University of Cambridge, 19 JJ Thomson Avenue, Cambridge CB3 0HE, UK\\%
$^{3}$Department of Physics, The University of Texas at Dallas, Richardson, Texas 75080, USA\\%
$^{4}$Department of Physics and Astronomy, York University, 4700 Keele Street, Toronto, ON M3J 1P3, Canada\\%
$^{5}$INAF — Arcetri Astrophysical Observatory, Largo E. Fermi 5, I-50125, Florence, Italy\\%
$^{6}$Center for Astrophysics | Harvard \& Smithsonian, 60 Garden Street, Cambridge, MA 02138, USA\\%
$^{7}$Department of Astronomy, Tsinghua University, Beijing 100084, People’s Republic of China\\%
$^{8}$Department of Physics and MIT Kavli Institute for Astrophysics and Space Research, 77 Massachusetts Avenue, Cambridge, MA 02139, USA%
}

\date{Accepted XXX. Received YYY; in original form ZZZ}

\pubyear{2024}

\begin{document}
\label{firstpage}
\pagerange{\pageref{firstpage}--\pageref{lastpage}}
\maketitle

\begin{abstract}
We present an analysis of the diffuse ionised gas (DIG) in a high-resolution simulation of an isolated Milky Way-like galaxy, incorporating on-the-fly radiative transfer and non-equilibrium thermochemistry. We utilise the Monte-Carlo radiative transfer code \textsc{colt} to self-consistently obtain ionisation states and line emission in post-processing. We find a clear bimodal distribution in the electron densities of ionised gas ($n_{\rm e}$), allowing us to define a threshold of $n_{\rm e}=10\,\mathrm{cm}^{-3}$ to differentiate DIG from \HII regions. The DIG is primarily ionised by stars aged 5\,--\,25\,Myr, which become exposed directly to low-density gas after \HII regions have been cleared. Leakage from recently formed stars ($<5$\,Myr) is only moderately important for DIG ionisation. We forward model local observations and validate our simulated DIG against observed line ratios in [\ion{S}{II}]/H$\alpha$, [\ion{N}{II}]/H$\alpha$, [\ion{O}{I}]/H$\alpha$, and [\ion{O}{III}]/H$\beta$ against $\Sigma_{\rm H\alpha}$. The mock observations not only reproduce observed correlations, but also demonstrate that such trends are related to an increasing temperature and hardening ionising radiation field with decreasing $n_{\rm e}$. The hardening of radiation within the DIG is caused by the gradual transition of the dominant ionising source with decreasing $n_{\rm e}$ from 0\,Myr to 25\,Myr stars, which have progressively harder intrinsic ionising spectra primarily due to the extended Wolf-Rayet phase caused by binary interactions. Consequently, the DIG line ratio trends can be attributed to ongoing star formation, rather than secondary ionisation sources, and therefore present a potent test for stellar feedback and stellar population models.
\end{abstract}

\begin{keywords}
radiative transfer -- galaxies: ISM -- ISM: structure -- ISM: lines and bands -- HII regions -- galaxies: stellar content
\end{keywords}



\section{Introduction}
\label{sec:Introduction}

The interstellar medium (ISM) of star-forming galaxies is both cradle and grave for the generations of stars embedded within. 
Throughout their lives, these stars batter the ISM with stellar winds, supernovae, and a flood of photons which leave it chemically enriched and injected with energy \citep{Girichidis:2020aa,Vogelsberger:2020aa}. 
While stellar populations shape the ISM, they are also shaped by it. The state of the ISM regulates the rate at which new stars can form, creating a coupling between the history and future of star formation in a galaxy \citep{Agertz:2013aa, Vogelsberger:2014aa, Vogelsberger:2014ab, Chevance:2020aa}.
Understanding how this coupling gives rise to a feedback cycle and impacts the properties of galaxies is a fundamental question in the study of galaxy evolution.

Much of our understanding in this area has been driven by observations of ionised gas in the ISM, which uniquely traces energetic processes and is conveniently bright in optical and UV emission lines \citep{Maiolino:2019aa}. 
The study of \HII regions has proved particularly fruitful.
Through the combination of \HII region observations and modelling \citep[e.g.][]{Ferland:2017aa}, we have been able to quantify the impact of feedback from young stars \citep{Lopez:2014aa,Barnes:2020aa,McLeod:2021aa,Della-Bruna:2022aa}, to identify the presence of AGN \citep{Baldwin:1981aa,Kewley:2001aa,Kewley:2006aa}, and to measure the most important properties of galaxies, such as star-formation rates \citep[SFRs;][]{Kennicutt:1983aa,Kennicutt:2012aa}, gas-phase metallicities \citep{Tremonti:2004aa,Maiolino:2019aa}, and electron densities \citep{Kewley:2019ab}.

However, a significant fraction of the ionised gas within the ISM of star forming galaxies is in the form of diffuse ionised gas (DIG), which is lower density than a typical \HII region and $\sim2000$\,K hotter \citep{Madsen:2006aa}. It is observed in edge-on galaxies as extraplanar gas extending on kpc scales above the thin disk and in face-on galaxies as lower surface brightness regions \citep{Levy:2019aa}.
The DIG contributes around half of the H$\alpha$ emission in star-forming galaxies, which implies that the DIG contributes substantially to the overall energy budget of the ionised phase of the ISM \citep{Oey:2007aa, Tacchella:2022aa}. 
The most obvious candidate for the source of DIG ionisation is leaking radiation from \HII regions, as the DIG is often observed to be spatially associated with \HII regions \citep{Levy:2019aa, Belfiore:2022aa}, however this has not been a universally accepted conclusion \citep{Seon:2009aa}.

To further complicate the picture, observations of metal line emission in the DIG have found that the line ratios [\ion{N}{II}]$\lambda$6585~\AA/H$\alpha$, [\ion{S}{II}]$\lambda$6718,33~\AA~/~H$\alpha$, [\ion{O}{I}]$\lambda$6302~\AA~/~H$\alpha$, and [\ion{O}{III}]$\lambda$5008~\AA~/~H$\beta$ increase with decreasing surface brightness, implying that they increase as ionised gas becomes more diffuse \citep{Hill:2014aa}. 
Photoionisation modelling can explain the [\ion{N}{II}]/H$\alpha$, [\ion{S}{II}]/H$\alpha$, and [\ion{O}{I}]/H$\alpha$ trends as a result of radiation leaking from an \HII region into and ionising low-density gas \citep{Belfiore:2022aa}.

However, the [\ion{O}{III}]/H$\beta$ trend is more difficult to explain. The ionisation energy of \OII is 35.1\,eV, which would imply that the radiation field becomes significantly harder as gas becomes more diffuse.
While radiation from \HII regions hardens as it leaks due to the photoionisation cross section of \HI decreasing with energy, this effect has so far not been sufficient to reproduce the observed line ratios in photoionisation models \citep{Wood:2004aa,Barnes:2015aa}.
Alternatively, additional sources of heating in the DIG may be sufficient to shift the emissivity ratio of [\ion{O}{III}]/H$\beta$ enough to reproduce the observed line ratios \citep{Rand:1998aa}. Potential mechanisms for this heating include cosmic rays \citep{Vandenbroucke:2018aa}, photoelectric heating of dust grains \citep{Reynolds:1992aa}, or turbulence \citep{Binette:2009aa}.

A secondary hard ionisation source in the DIG could cause an increased abundance of \OIII and therefore explain the observed line ratios. 
The most promising secondary ionisation source is the contribution of stars with masses $0.8-8~\Msun$ in evolutionary phases after the asymptotic giant branch stage (post-AGB), also called hot low-mass evolved stars (HOLMES). These stars have a hard ionising spectrum, and could provide a significant contribution to the budget of high-energy photons while still maintaining a low overall contribution to the total emission of ionising photons \citep{Flores-Fajardo:2011aa}.
The importance of old stars in ionising the ISM of elliptical and lenticular galaxies has been well studied \citep{Sarzi:2010aa, Singh:2013aa}, but their contribution has also been invoked by \citet{Zhang:2017aa} to explain the line ratio trends in star-forming galaxies observed in the MaNGA survey \citep{Bundy:2015aa}.

The recent study by \citet{Belfiore:2022aa} examines ionisation sources within the DIG by utilising emission line maps from face-on, star-forming galaxies in the PHANGS-MUSE survey \citep{Emsellem:2022aa}. This work offers the most detailed perspective of the DIG to date, enabling us to resolve the ISM in emission at scales of tens of parsecs. Furthermore, this work also supports the idea that the observed line ratios are significantly influenced by HOLMES.

These conclusions regarding the DIG conditions are based on relatively simple photoionisation models. While such models are highly effective for gaining insights about \HII regions, which are localized and comparatively isolated systems, they are not entirely adequate for capturing the full complexity of the DIG. Given that the DIG represents a galaxy-scale phenomenon, it necessitates a more comprehensive approach. This involves self-consistent modelling of heating, ionisation, and line emission across the entire scale of a galaxy.

Simulating a realistic, resolved ISM poses significant challenges \citep{Hopkins:2014aa, Naab:2017aa}. It requires combining high spatial and mass resolution with computationally demanding physics on-the-fly, including multi-wavelength radiative transfer, dust formation and destruction, and sophisticated stellar feedback subgrid recipes \citep{Kannan:2020aa}.
Furthermore, accurately calculating line emission requires knowledge of the relative abundances of ionisation states. However, on-the-fly calculations of ionisation states for metal ions remain prohibitively expensive for general adoption \citep{Katz:2022ad}. Additionally, transient radiative transfer phenomena can introduce numerical artefacts that can lead to unrealistic emission line fluxes \citep{Smith:2022aa,Deng:2024aa}. Consequently, in many cases post-processing is essential to address the radiative transfer of ionising radiation and calculate self-consistent ionic abundances accurately.
Lastly, a careful treatment of the radiative transfer of the emission lines themselves is crucial, including dust scattering and absorption to generate realistic observed emission line maps for direct comparison with observations.

Such comprehensive modelling has so far only been achieved for tall box theoretical studies of the DIG \citep[e.g.][]{Vandenbroucke:2018aa, Vandenbroucke:2019aa}. These contextual constraints have limited our understanding of important aspects of stellar feedback, such as the leakage of radiation from \HII regions \citep{Chevance:2022aa}. Overall, this has resulted in a reliance on strong line diagnostics, which are susceptible to contamination by the \citep{Zhang:2017aa}, leaving us without precise guidance in our interpretations.

In this work, we aim to address the theoretical shortcomings by employing the recently developed capability for metal ionisation and line emission in the Monte Carlo radiative transfer (MCRT) Cosmic Ly$\alpha$ Transfer code \citep[\textsc{colt};][]{Smith:2015aa, Smith:2019aa, Smith:2022aa} to self-consistently model the resolved optical metal line emission from a simulated Milky Way-like (MW-like) galaxy. Our approach builds upon the work of \citet{Tacchella:2022aa}, wherein this model was demonstrated to effectively reproduce realistic Balmer line emission from the DIG.

Utilizing this methodology, our study is focused on addressing the following questions:
\begin{enumerate}
\item Do the observed trends in the DIG align with those predicted by a purely theoretical definition of the DIG?
\item Are the variations in line ratios within the DIG primarily influenced by changes in ionic abundances, or by line emissivity due to temperature variations?
\item What are the respective contributions of \HII region leakage, old stars, and shocks in generating these line ratios?
\end{enumerate}

In Section~\ref{sec:Methods}, we elaborate on the methodologies employed for simulating the MW-like galaxy and the subsequent post-processing of emission lines.
In Section~\ref{sec:ComparisontoObservations}, we compare our mock emission-line observations to surface brightness profiles and DIG line ratio trends from the MaNGA and PHANGS-MUSE surveys, respectively.
In Section~\ref{sec:PhysicsoftheDIG}, we discuss the underlying mechanisms contributing to the observed line ratio trends in the DIG with regard to the transport of ionising photons and the temperature variations within the DIG.
In Section~\ref{sec:Discussion} we discuss our results in the context of future observational studies of the DIG. We also consider how the ionisation of the DIG is dependant on early stellar feedback and the ionising SEDs of young stars, in particular the inclusion of binary stars.
We conclude in Section~\ref{sec:Conclusions}, where present our main results.

Throughout this paper, we will employ the following shorthand notations for convenience: [\ion{N}{II}] denotes [\ion{N}{II}]$\lambda$6585~\AA, [\ion{S}{II}] represents [\ion{S}{II}]$\lambda$6718,33~\AA, [\ion{O}{I}] refers to [\ion{O}{I}]$\lambda$6302~\AA, and [\ion{O}{III}] denotes [\ion{O}{III}]$\lambda$5008~\AA.

\section{Methods}
\label{sec:Methods}

In this section we describe the methods employed to accurately model the ISM and create mock emission line observations.
We begin in Section~\ref{sec:IsolatedMWSimulation} by describing the simulation of a MW-like galaxy.
In Section~\ref{sec:RadiativeTransferofIonisingPhotons} and Section~\ref{sec:RadiativeTransferofEmissionLinePhotons}, we describe the post-processing treatments of ionising and emission line photons, respectively.

\subsection{Isolated MW simulation}
\label{sec:IsolatedMWSimulation}

\begin{figure*} 
\centering
	\includegraphics[width=\textwidth]{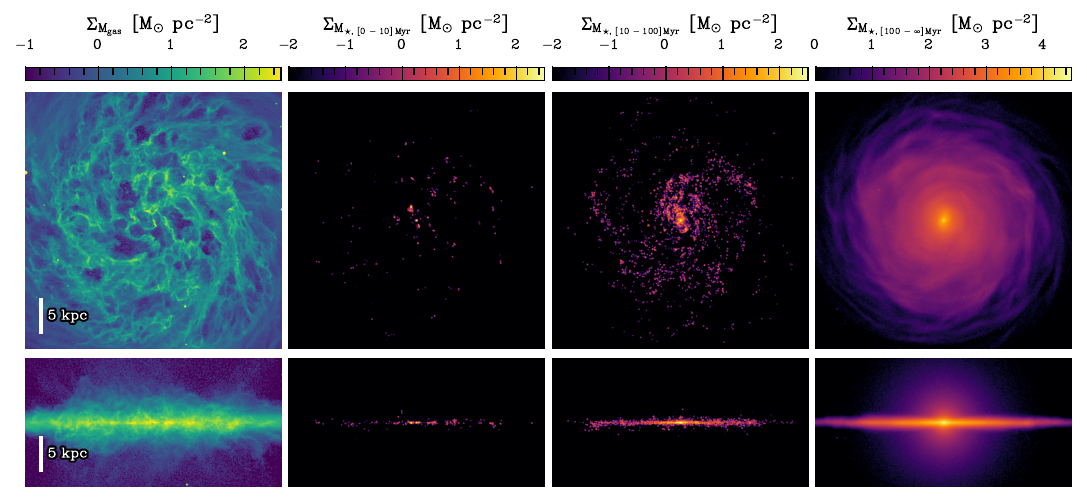}
    \caption{Projected properties of the MW-like simulation at 713\,Myr, including 50\,pc smoothed surface density maps displayed from left to right of gas mass, stellar mass in young stars ([0\,--\,10]\,Myr), stellar mass in intermediate-aged stars ([10\,--\,100]\,Myr), and stellar mass in older stars ([100\,--\,$\infty$]\,Myr). The maps highlight the stochastic nature of star formation, which is especially evident in the strongly non-uniform distribution of recent star formation observed over 10\,Myr timescales.}
    \label{fig:mw_properties}
\end{figure*}

In this study, we utilize a high-resolution simulation of a galaxy designed to resemble the Milky Way, characterized by a halo mass of $M_{\mathrm{halo}} = 1.5 \times 10^{12}\,\Msun$, stellar bulge and disk mass of $M_{\mathrm{stars}} = 6.2 \times 10^{10}\,\Msun$, and gas mass of $M_{\mathrm{gas}} = 9 \times 10^9\,\Msun$. This simulation was previously presented by \citet{Kannan:2020aa} and \citet{Kannan:2021aa}, but we summarise the most relevant details below. 
The simulation employs \textsc{arepo-rt} \citep{Kannan:2019aa}, which is an extension of the moving mesh hydrodynamic code \textsc{arepo} \citep{Springel:2010aa,Weinberger:2020aa}\footnote{The public code and its documentation are accessible at \href{https://arepo-code.org}{\texttt{arepo-code.org}}.}. The sub-grid models, which are crucial for capturing star formation and feedback processes, are described in \citet{Marinacci:2019aa} and \citet{Kannan:2020aa}. Important projected quantities derived from the simulation are illustrated in Fig.~\ref{fig:mw_properties}.

The simulation aims to self-consistently resolve the ISM on scales of $\sim10$\,pc. This is achieved by integrating on-the-fly radiation hydrodynamics (RHD) with non-equilibrium thermochemisty, alongside mechanisms for dust formation and destruction, and employing the Stars and MUltiphase Gas in GaLaxiEs (SMUGGLE) feedback model \citep{Marinacci:2019aa}.
The simulation is run inside a 600\,kpc box and is composed of a static dark matter halo, a bulge, a stellar disk, and a gaseous disk \citep{Hernquist:1993aa, Springel:2005aa}.
The dark matter halo and bulge adopt a Hernquist profile with a 1\,kpc scale length \citep{Hernquist:1990aa}.
The stellar disk follows a radial exponential profile with an effective radius $R_{\mathrm{eff}} = 2\,\mathrm{kpc}$ and a vertical $\mathrm{sech}^2$ profile with a 300\,pc scale height, and the initial stars are assigned an age of 5\,Gyr.
The gas disk also follows a radial exponential profile with the same effective radius as the stellar disk. It is vertically distributed in hydrostatic equilibrium, starting with an initial temperature of $10^4\,\mathrm{K}$ and a metallicity of $1\,\Zsun$. The idealised nature of the simulation and lack of cosmological environment could lead to unrealistic gas metallicities, so the creation of new metals is turned off. This has the benefit of isolating our investigation of the ionisation and distribution of the DIG from the effects of metallicity gradients and metallicity evolution between snapshots.
The simulation runs for 1\,Gyr and, following an initial starburst, reaches a quasi-equilibrium state around $\sim200\,\mathrm{Myr}$.

The thermochemical network tracks the ionic species \ion{H}{$_{2}$}, \ion{H}{I}, \ion{H}{II}, \ion{He}{I}, \ion{He}{II}, and \ion{He}{III}.
The RHD implementation evolves the radiation field in distinct energy bins: infrared (0.1\,--\,1\,eV), optical (1\,--\,11.2\,eV), Lyman-Werner (11.2\,--\,13.6\,eV), \ion{H}{I} ionising (13.6\,--\,24.6\,eV), \ion{He}{I} ionising (24.6\,--\,54.4\,eV), and \ion{He}{II} ionising (54.4\,--\,$\infty$\,eV).
These energy ranges are chosen to differentiate photons based on their relevance to various physical processes, such as the ionisation of different species, photodissociation of \ion{H}{$_{2}$}, radiation pressure, photoelectric heating, and photoheating.

Tracking the abundances of molecular hydrogen and the ionisation of atomic hydrogen and helium also allows for self-consistent modelling of primordial cooling due to these species alongside corresponding heating and cooling rates.
However, metal ions are not directly tracked within the model. Instead, their cooling rate is estimated using a look-up table derived from CLOUDY simulations \citep{Vogelsberger:2013aa}.
Dust heating processes are also accounted for, including gas-dust collisional energy exchange and photoelectric heating, which occurs when photons with energies between 5.8\,--\,11.2\,eV eject electrons from dust grains.
Additionally, cosmic ray ionisation and heating are included by assuming canonical MW ionisation and heating rates.
The simulation allows gas to cool down to a temperature floor of $10$\,K.

Dust is dynamically coupled to the gas, and can be created and destroyed through a self-consistent dust model \citep{McKinnon:2017aa}. Dust is produced through mass return from Type II Supernovae (SNII), Type Ia Supernovae (SNIa), and asymptotic giant branch (AGB) stars. The dust mass increases due to gas-phase metals in the ISM depositing onto grains. Dust is destroyed via shocks from supernova (SN) remnants or through thermal sputtering. This process results in a complex distribution of dust that cannot be replicated using simple scaling relations in post-processing, such as with a dust-metallicity relation \citep{McKinnon:2016aa,McKinnon:2018aa}.

Stars are formed probabilistically from cold, self-gravitating gas with a minimum density of $n=10^{3}\,\mathrm{cm}^{-3}$. Three primary mechanisms of stellar feedback are included: radiative feedback, stellar winds from AGB stars and young OB stars, and SN feedback. 
Radiative feedback is accurately modelled through the RT scheme, representing star particles as radiation sources and utilizing the spectral energy distribution (SED) from the stellar population synthesis model presented in \citet{Bruzual:2003aa}. 
Stellar winds and SN feedback are injected using the SMUGGLE subgrid prescriptions.

\subsection{Radiative transfer of ionising photons}
\label{sec:RadiativeTransferofIonisingPhotons}

To accurately predict line emission from the simulation, we must know the relative ionic abundances in each gas cell. The isolated MW-like simulation tracks the relative abundances of \ion{H}{$_{2}$}, \ion{H}{I}, \ion{H}{II}, \ion{He}{I}, \ion{He}{II}, \ion{He}{III}, which in theory makes it possible for us to directly calculate the Balmer line emission. However, using the on-the-fly ionisation state abundances for Balmer line emission estimation can result in observed luminosities that are an order of magnitude too high. This discrepancy is primarily due to very young \HII regions within the simulation lacking a fully resolved substructure in terms of density and temperature, as discussed in \citet{Smith:2022aa}.

Consequently, it is necessary to recompute the ionisation states of \ion{H}{} and \ion{He}{} using a post-processing approach. Moreover, considering our interest in spatially- and spectrally-resolved line emission from metals, a post-processing approach is also essential because the relative abundances of metal ions are not tracked on-the-fly.

We employ \textsc{colt} to perform MCRT of ionising radiation and thereby obtain ionisation states for each simulated gas cell. The primary goal of our post-processing method is to enhance and supplement the physics modelled on-the-fly, while preserving the accuracy of gas properties such as density and internal energy that were reliably simulated during the on-the-fly calculations. This approach is particularly advantageous for studying the DIG because we are interested in understanding how temperature and emissivity variations affect the observed line ratios. Achieving a self-consistent analysis would be challenging if we modified the gas temperature during post-processing by considering only photoheating. Retaining the original temperature data also provides a more focused exploration, and makes our study complementary to previous works \citep[e.g.][]{Tacchella:2022aa,Yang:2023aa}.

We now briefly describe the ionisation equilibrium calculations. Ionising photons are primarily emitted by young, massive stars, while the star particles in the simulation represent initial mass function (IMF) averaged stellar populations. These sources are characterized by SEDs derived from the Binary Population and Spectral Synthesis (BPASS) model \citep[v2.2.1;][]{Eldridge:2009aa, Eldridge:2017aa}\footnote{Further information on BPASS can be found on the project website at \href{https://bpass.auckland.ac.nz}{\texttt{bpass.auckland.ac.nz}}.}.
To ensure the stellar emission is well-sampled with a sufficient number of directions from less luminous sources, we apply a luminosity boosting technique that biases the MCRT source probability distribution and corresponding weight assignments by a power law proportional to $\propto L^{3/4}$ for sampling photon packets. This allows for a more balanced and robust sampling across the entire spectrum of stellar sources.

In our procedure, the sampling of ionising photons is organised into disjoint energy bins.
The specific boundaries and number of these bins are determined based on the ions that are being tracked in post-processing. Specifically, we add all ions with ionisation thresholds below 100\,eV and one higher level implicitly from the H, He, C, N, O, Ne, Mg, Si, S, and Fe atomic species. We include each of these thresholds as energy bin edges, and further subdivide bins by enforcing a minimum bin width of 0.05\,dex, for a total of 90 bins. This relatively high-resolution multi-wavelength strategy was chosen to guarantee that emitted photons have the requisite energy for ionisation. Coarse binning can result in false ionisation due to under-resolving sharp opacity features. High spectral resolution is also crucial for properly keeping track of the hardness of the radiation field. Beyond this, we apply a frequency biasing scheme of $\nu^2$ when sampling the bin to better probe the high-energy end of the SED. Finally, we uniformly suppress low-energy sampling to ensure that 80 per cent of photon packets are hydrogen ionising with energies above 13.6\,eV.

When an ionising photon packet is sampled from a star in our simulation, it is launched into the ISM with an isotropic randomly assigned direction, starting from the position of the star. The gas cells within the simulation are represented by a Voronoi tessellation, and we employ ray-tracing techniques to accurately propagate the photon through the ISM, as introduced to \textsc{colt} by \citet{Smith:2017aa}. The dust distribution is taken directly from the simulation, and the dust opacity, albedo, and anisotropic scattering parameters are taken from the Milky Way dust model of \citet{Weingartner:2001aa}. We thus maintain a faithful representation of the \textsc{arepo} geometry and simulated gaseous and dusty components of the ISM.

Each photon packet in the simulation is assigned an initial weight.
As it traverses through the ISM, this photon packet can undergo various interactions, including attenuation by dust, scattering by dust, or photoionisation of atoms and ions in the gas. Absorption processes are modelled continuously, with the weight of the photon packet decreasing exponentially by $e^{-\tau_\text{a}}$ for each distance segment $\Delta\ell$ it travels. Here, $\tau_\text{a}$ is the total pure absorption optical depth along that distance in a particular energy bin. Specifically, this is given as the sum of the absorption coefficients for dust and gas, i.e. $k_\text{a} \equiv \text{d}\tau_\text{a}/\text{d}\ell = (1 - A) \kappa_\text{d} \mathcal{D} \rho + \sum_x n_x \sigma_x$, where $A$ denotes the scattering albedo, $\kappa_\text{d}$ the dust opacity, $\mathcal{D}$ the dust-to-gas ratio, $\rho$ the gas density, $n_x$ the number densities of each ionic species $x$, and $\sigma_x$ the photoionisation cross-sections of the ion in the given band \citep[for details see][]{Smith:2022aa}.

The remaining MCRT procedures within the simulation are fairly standard. Dust scattering is accounted for by determining a scattering distance using the exponential distribution, with a scattering coefficent given by $k_\text{s} \equiv \text{d}\tau_\text{s}/\text{d}\ell = A \kappa_\text{d} \mathcal{D} \rho$ for the relevant energy band.
Photon trajectories are terminated when its weight becomes negligible, which is set to be when it falls below $10^{-14}$ times the total ionising photon emission rate from stars.
In addition to the stellar photoionisation already described, the ionisation equilibrium solver also accounts for recombinations, collisional ionisations, charge exchange reactions, and a meta-galactic UV background, as modelled by \citet{Faucher-Giguere:2009aa}, and incorporates self-shielding following the approach of \citet{Rahmati:2013aa}. Finally, as the transport opacities depend on the ionisation states, \textsc{colt} employs iteration between MCRT (with $10^8$ photon packets) and abundance updates until the global number of recombinations is converged to within $0.1$ per cent.

A significant benefit of the MCRT approach for calculating ionisation states is the ability to keep track of each photon packet individually. This allows us to create aggregate statistics about these packets, such as their mean free path or the fraction absorbed by dust. This level of detail is particularly valuable in our study of the DIG, where we are interested in a comprehensive understanding of the sources and sinks of ionising radiation.

\subsection{Radiative transfer of emission line photons}
\label{sec:RadiativeTransferofEmissionLinePhotons}

\begin{figure*} 
\centering
	\includegraphics[width=\textwidth]{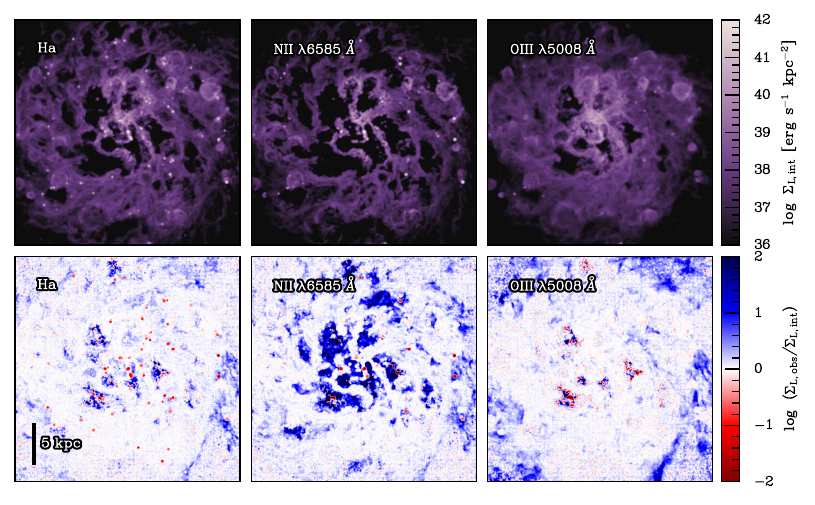}
    \caption{H$\alpha$, [\ion{N}{II}], and [\ion{O}{III}] maps for our MW-like simulation at 713\,Myr. Maps have been smoothed by 100\,pc to improve visibility. \textit{Top:} The intrinsic surface brightness of each emission line, $\Sigma_{\mathrm{L, int}}$. \textit{Bottom:} The ratio of observed-to-intrinsic surface brightness in each pixel. The compact, bright regions around recent star formation are heavily attenuated due to the high concentration of dust. On the other hand, low surface brightness regions are boosted due to dust scattering photons into the line of sight. }
    \label{fig:metalemission}
\end{figure*}

After establishing the ionic abundances for each cell in the simulation, our next step is to calculate the line emission that arises from this ionisation state distribution. Fundamentally, this process still involves solving for the radiative transfer of photons through the ISM. However, a crucial distinction here is that the source of these photons is diffuse in nature; i.e. we are now sampling emission-line photons from gas cells rather than ionising photons originating from stars. This shift to line radiative transfer means that the included physics is more focused; e.g. we employ continuous frequency tracking with Doppler-shifting between comoving gas frames for accurate spatially- and spectrally-resolved synthetic observations.

In the process of performing MCRT for emission lines and generating emission-line maps, we again utilise the \textsc{colt} code. Examples of the resulting surface brightness images for the H$\alpha$, [\ion{N}{II}], and [\ion{O}{III}] lines are illustrated in Fig.~\ref{fig:metalemission}.

To accurately determine the luminosity of the gas cells for a given emission line, it is essential to account for the two primary mechanisms through which emission-line photons are emitted: collisional excitation and recombination.
Collisional excitation occurs when an electron collides with an atom, transferring energy that excites another bound electron to a higher energy state. The subsequent radiative deexcitation emits a line photon. Recombination, on the other hand, involves the capture of a free electron by an ionised atom, resulting in the release of line photons as the system cascades to lower energy states.
For most emission lines, one of these channels dominates the other to such an extent that lines are often referred to as collisionally excited lines (CELs) or recombination lines (RLs). This is certainly the case for the emission lines we examine in this work: H$\alpha$, H$\beta$, [\ion{N}{II}], [\ion{S}{II}], [\ion{O}{I}], and [\ion{O}{III}]. For the RLs H$\alpha$ and H$\beta$, we consider the dominant recombination channel and the minor collisional excitation channel. The metal lines in this work are all CELs and we only consider the dominant collisional channel for them.
Still, information from multiple emission lines is highly complementary for understanding the conditions within the ISM and DIG of galaxies, owing to the specific density, temperature, and metallicity environmental dependencies.

The RL and CEL channels pose distinct challenges in our modelling approach. Let us first consider a recombination line, which has its luminosity in a particular volume given by
\begin{equation}
  L_X^\mathrm{rec} = h \nu_X \int P_{\mathrm{B},X}(T,n_{\rm e}) \alpha_\mathrm{B}(T)\,n_{\rm e} n_\text{ion}\,\text{d}V \, ,
\end{equation}
where $X$ denotes the line, the energy at line centre is $h\nu_X$, and $n_{\rm e}$ and $n_\text{ion}$ are the number densities for free electrons and ions, respectively. $\alpha_\mathrm{B}$ is the case B recombination coefficient, and $P_{\mathrm{B},X}$ is the probability for that emission line photon to be emitted per recombination event. The coefficient $\alpha_\mathrm{B}$ used here is the same as that employed in the ionisation equilibrium solver; i.e. from \citet{Hui:1997aa}.
At high densities and low temperatures the equilibrium solution essentially reduces to photoionisations balancing recombinations. Therefore, the overall RL luminosities in our simulation are closely tied to the previously modelled ionising radiative transfer. This link is crucial for ensuring the accuracy and consistency of RL predictions.

When calculating the luminosity of RLs in cold, dense, partially ionised gas, our model encounters specific challenges.
Such gas states, while non-physical, occasionally arise through transient numerical phenomena in the simulation, particularly in scenarios where a star has recently formed but has not yet had sufficient time to heat its birth cloud.
The main issue here is not with the total luminosity, which for RLs is effectively constrained by the total number of photoionisations, but rather with the emission probability per recombination event.
For example, this can result in unrealistic Balmer decrements.
To address this, we adopt a temperature floor of 7000\,K specifically for calculating $P_{\mathrm{B},X}$. We emphasise that this floor is only applied in the context of calculating emission probabilities, and not for determining the total recombination rate, which remains entirely self-consistent with the ionisation equilibrium step.

CELs present a more complex modelling challenge because their relationship to photoionisation is not as direct as that of RLs. The luminosity for a cell in a particular CEL is given by
\begin{equation}
  L_X^\mathrm{col} = h \nu_X \int q_{\mathrm{col},X}(T)\,n_{\rm e} n_\text{ion}\,\text{d}V \, ,
\end{equation}
where $q_{\mathrm{col},X}$ is the collisional excitation rate coefficient.
This coefficient has an exponential dependence on temperature, meaning that an excessively high temperature in a partially ionised cell could rapidly lead to unrealistically high luminosities. Managing this temperature dependency is crucial for accurately simulating CELs and ensuring that the emission characteristics of the gas cells reflect realistic physical conditions. In the present work, we have chosen to only directly address this for hydrogen lines \citep[as outlined in][]{Smith:2022aa}. The metal lines considered herein are also likely to be affected by temperature sensitivity and under-resolved partial ionisation substructure. However, we expect this to have minimal implications for the primary goals of this paper of understanding DIG emission, especially as this particular simulation achieves a relatively high mass resolution ($m_\text{gas} \sim 10^3\,\Msun$). Nonetheless, we have been actively developing robust CEL correction strategies and refinement schemes for other applications.

\section{Comparison to Observations}
\label{sec:ComparisontoObservations}

In this section, we show that our mock observed line ratios match observations of local star-forming galaxies. 
In Section~\ref{sec:RadialProfilesfromSDSSMaNGA}, we show that the surface brightness profiles of our simulated emission lines match those in the SDSS/MaNGA survey \citep{Bundy:2015aa}.
In Section~\ref{sec:DIGLineRatiosfromMUSE}, we show that when treating our mock observations as an observer would when analysing the DIG, we recover the the observed line ratio trends of galaxies in the PHANGS-MUSE survey \citep{Emsellem:2022aa}.

\subsection{Radial profiles from SDSS/MaNGA}
\label{sec:RadialProfilesfromSDSSMaNGA}

The SDSS/MaNGA survey \citep{Bundy:2015aa} provides spatially-resolved emission line maps for a large sample of galaxies. By comparing the surface brightness profiles of MaNGA galaxies to those generated from our simulation with \textsc{colt}, we can confirm that we are producing realistic mock observations. 

\begin{figure}
    \centering
	\includegraphics[width=\columnwidth]{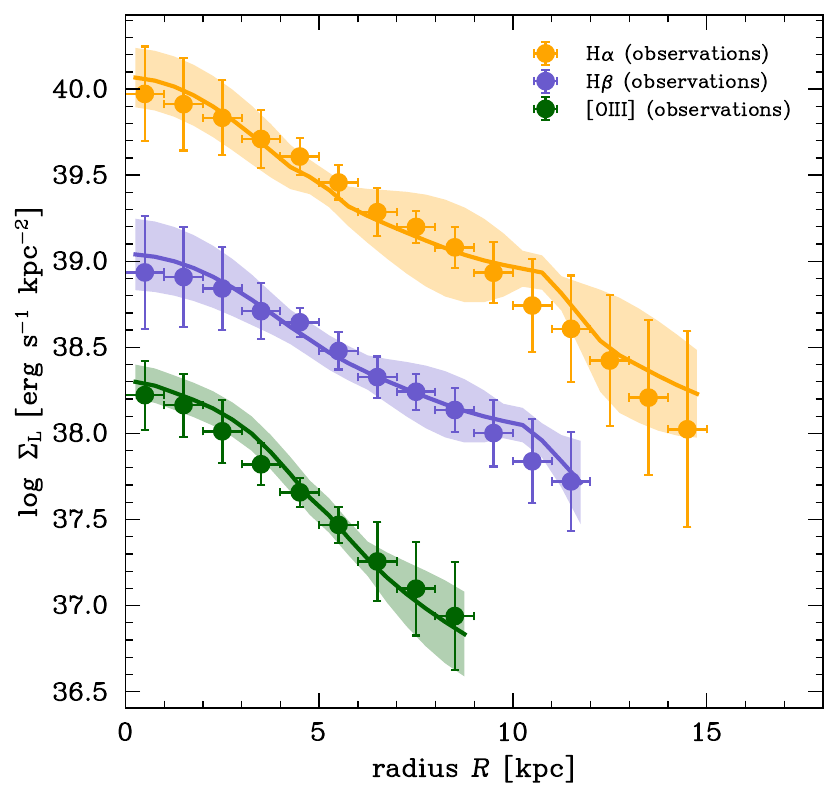}
    \caption{Comparison of median radial surface brightness profiles of H$\alpha$, H$\beta$, and [\ion{O}{III}] between the sample of MW-like galaxies from MaNGA and four snapshots of our simulated MW galaxy. The errorbars and shaded regions represent the 16--84$^\text{th}$ percentile scatter for the MaNGA sample and sampled simulation snapshots respectively. The H$\alpha$, H$\beta$, and [\ion{O}{III}] profiles are normalised to a total luminosity of $10^{42}\,\mathrm{erg}\,\mathrm{s}^{-1}$, $10^{41}\,\mathrm{erg}\,\mathrm{s}^{-1}$, and $10^{40}\,\mathrm{erg}\,\mathrm{s}^{-1}$, respectively.
    Our MW simulation modelling successfully produces realistic surface brightness profiles.}
    \label{fig:mangacomparison}
\end{figure}

Previous comparisons between the H$\alpha$ and H$\beta$ surface brightness profiles from our MW-like simulation and the MaNGA observations have shown good agreement \citep{Tacchella:2022aa}. In this study, we extend this comparison to include metal emission lines, using the same subset of 52 MaNGA-observed galaxies that resemble the MW.
Galaxies in the sample were selected based on specific criteria: elliptical Petrosian photometry stellar masses from the NASA-Sloan Atlas \citep{Blanton:2005aa,Blanton:2011aa} in the range $\log(M_{\star}/\mathrm{M}_{\odot})=10.6$--$10.9$ and SFRs calculated with H$\alpha$ to be between $\mathrm{SFR}/(\mathrm{M}_{\odot}\,\mathrm{yr}^{-1})=0.5$--$5.0$.
Galaxies were rejected from the sample if labelled with DAP datacube quality flags cautioning `do not use', if labelled with two or more `warning' flags, or if they were determined to be irregular following visual inspection of the H$\alpha$ map.

To account for the MaNGA survey resolution, characterised by a point spread function (PSF) with a full width at half-maximum (FWHM) of 2.5 arcsec (equivalent to 3.9\,kpc at the average sample galaxy distance of $z=0.078$), we applied a Gaussian smoothing kernel to our simulated emission line maps with a matching FWHM.

Fig.~\ref{fig:mangacomparison} shows a comparison of the H$\alpha$, H$\beta$, and [\ion{O}{III}] surface brightness profiles between the MW-like selected MaNGA galaxies and our simulations (based on four snapshots).
The surface brightness profiles from our simulation closely match MaNGA, which strengthens our confidence in the simulation's ability to accurately replicate line emission from a MW-like galaxy.

\subsection{Line ratios from MUSE}
\label{sec:DIGLineRatiosfromMUSE}

While the agreement of radial surface brightness profiles with MaNGA data helps give us confidence that our post-processing approach has produced physically reasonable emission line maps, the definitive test is to recreate the specific observational phenomena we are seeking to explain.
In particular, we want to reproduce the trends observed between line ratios and H$\alpha$ surface brightness ($\Sigma_{\rm H\alpha}$) in the DIG as reported in \citet{Belfiore:2022aa}.

\begin{figure*} 
\centering
	\includegraphics[width=\textwidth]{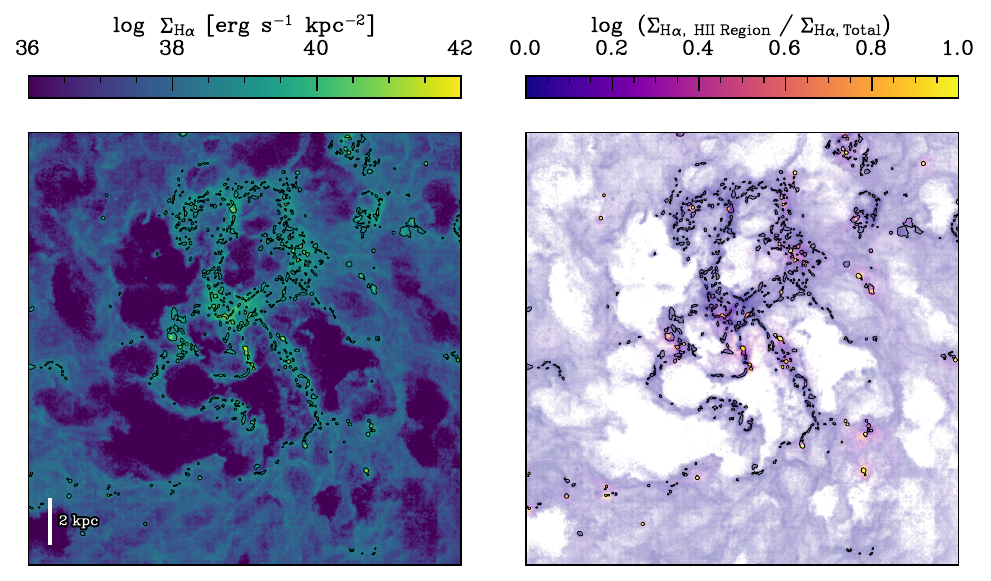}
    \caption{Observational selection of the DIG.
    \textit{Left:} A mock MUSE H$\alpha$ emission line map with black contours illustrating the \HII regions selected using the dendrogram method. \textit{Right:} The fraction of H$\alpha$ emission in a spaxel from gas with a density larger than $n_{\rm e}=10\,\mathrm{cm}^{-3}$, our theoretical demarcation between \HII regions and the DIG. The image transparency is based on the H$\alpha$ surface brightness. The dendrogram selection includes most spaxels which are dominated by the theoretically defined \HII regions, however it also selects a significant number of DIG dominated spaxels.}
    \label{fig:dendrogram}
\end{figure*}

\begin{figure*} 
\centering
	\includegraphics[width=\textwidth]{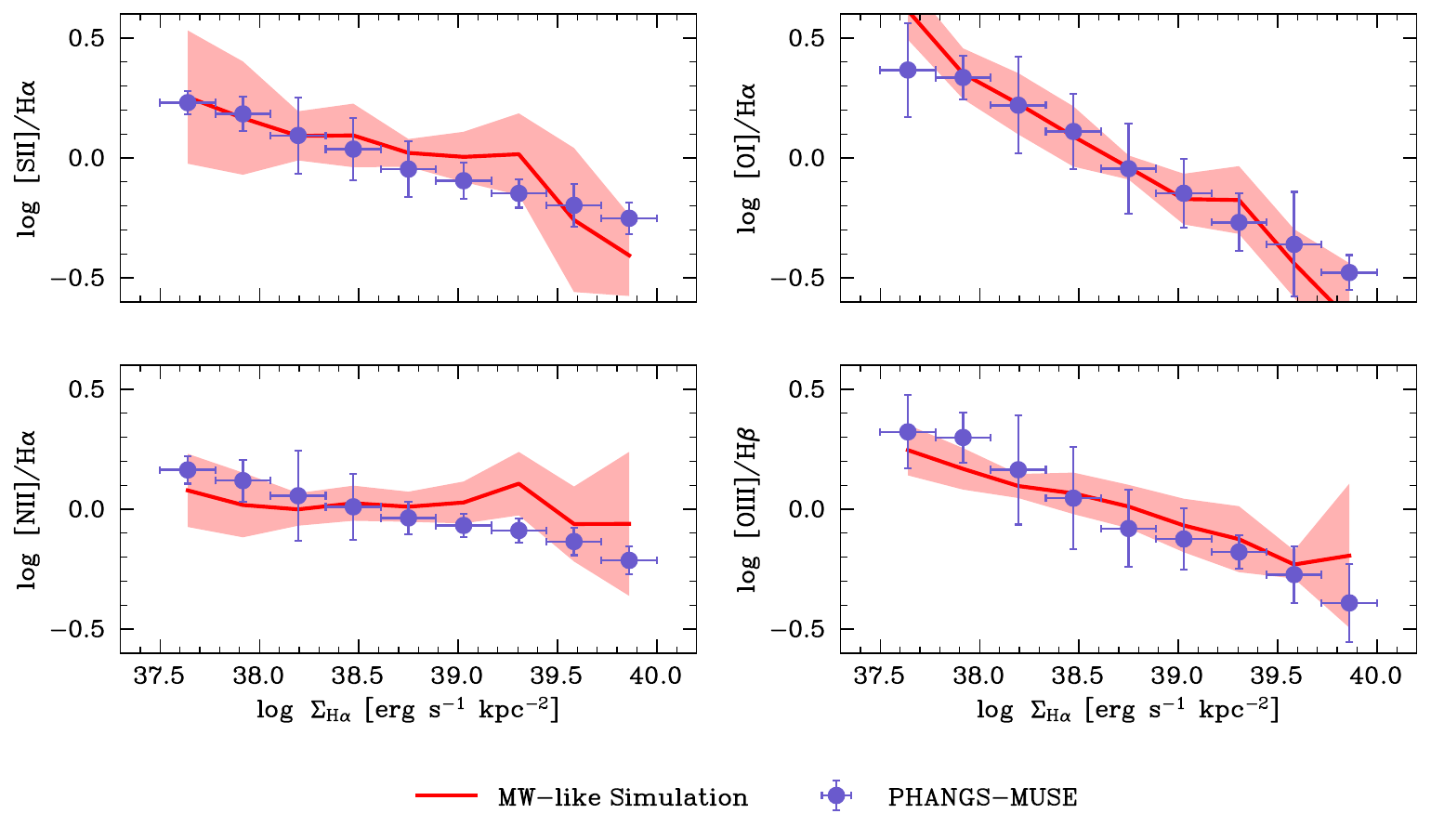}
    \caption{An observers view of the [\ion{S}{II}]/H$\alpha$, [\ion{N}{II}]/H$\alpha$, [\ion{O}{I}]/H$\alpha$, and [\ion{O}{III}]/H$\beta$ line ratios as functions of $\Sigma_{\rm H\alpha}$ in the DIG for MW-like PHANGS-MUSE observed galaxies \citep{Belfiore:2022aa} and our simulation. For both, each galaxy was divided into five radial bins and the DIG-dominated spaxels were selected using morphological methods. The line ratio profiles were calculated from these spaxels and normalised such that the average line ratio in each radial bin is unity, to facilitate direct comparison of line ratio trends across galaxies. The data points and trendline represent the median profiles (errorbars and shaded regions represent the 16--84$^\text{th}$ percentile scatter)for the PHANGS-MUSE galaxies and four simulation snapshots with five radial bins each. The trends of our simulation generally agree with those in observed MW-like galaxies, giving us confidence that our simulated DIG mirrors the physical conditions of the DIG in reality. While the [\ion{N}{II}]/H$\alpha$ ratio is consistent within the scatter, it appears flatter than observations, possibly because we do not capture the full variety of MW-like galaxies (see Section~\ref{sec:DIGLineRatiosfromMUSE}).}
    \label{fig:musecomparison}
\end{figure*}

To achieve this, we must carefully account for observational effects to generate mock MUSE emission line maps and then analyse them as an observer. In general we follow the methods from \citet{Belfiore:2022aa}.

We now outline the procedure. First, we employ \textsc{colt} to model observed H$\alpha$ emission line maps at 10\,pc resolution. We then apply a Gaussian PSF with a FWHM of 60\,pc, comparable to the PHANGS-MUSE sample of galaxies, and rebin to a resolution of 20\,pc, equivalent to the 0.2$''$ pixels for the PHANGS-MUSE sample. Finally, we introduce realistic noise levels by adding random Gaussian fluctuations to each spaxel with a mean of $2\times 10^{37}\,\mathrm{erg}\,\mathrm{s}^{-1}\,\mathrm{kpc}^{-2}$. This gives an average SNR of $\sim3$ in spaxels with $\Sigma_{\rm H\alpha}=6\times 10^{37}\,\mathrm{erg}\,\mathrm{s}^{-1}\,\mathrm{kpc}^{-2}$, which again matches the PHANGS-MUSE data \citep{Emsellem:2022aa}.

Having created our mock MUSE H$\alpha$ maps, we now apply observational techniques to differentiate \HII regions from the DIG. \citet{Belfiore:2022aa} used the \textsc{HIIPHOT} code to select their \HII regions via the spatial gradient of H$\alpha$. Dendrograms are another common way to select \HII regions in emission line maps, and we employ this technique as presented in \citet{McLeod:2021aa}. We expect this change to have little effect on the line ratio trends because \citet{Belfiore:2022aa} show that the trends are robust to a wide variety of \HII region selection criteria.

We use \textsc{astrodendro} on our mock H$\alpha$ maps to generate an initial dendrogram. We set the detection flux threshold to $2\times 10^{38}\,\mathrm{erg}\,\mathrm{s}^{-1}\,\mathrm{kpc}^{-2}$ and require a minimum of 9 pixels. This detection flux is conservative relative to the noise in our mock observations. As our maps are high resolution, dendrograms are prone to over-dividing clearly connected \HII regions into many small clusters. To remedy this, we apply the \textsc{scimes} package to our initial dendrogram \citep{Colombo:2015aa}. The \textsc{scimes} package is an unsupervised pattern recognition program that is routinely applied to observations of giant molecular clouds. By applying it to the dendrogram created with \textsc{astrodendro}, we are able to robustly combine the over-divided \HII regions. We also checked that varying the input parameters to \textsc{astrodendro} and \textsc{scimes} did not significantly impact the observed DIG line ratio trends.

Fig.~\ref{fig:dendrogram} shows the selected \HII regions overlaid on the mock MUSE H$\alpha$ emission line map. With our \HII regions selected, we can mask them out and identify the remaining spaxels as part of the DIG. To further increase the SNR, particularly for the fainter metal lines, we again rebin the DIG spaxels to 140\,pc resolution, equivalent to 1.4$''$ spaxels in the PHANGS-MUSE sample \citep{Belfiore:2022aa}. Finally, we utilised the \textsc{vorbin} package on the diffuse pixels to achieve a minimum H$\alpha$ SNR of 60. \textsc{vorbin} uses a Voronoi binning technique to combine lower S/N spaxels together to increase the SNR at the expense of spatial resolution. 

Using these binned spaxels, we plot [\ion{S}{II}]/H$\alpha$, [\ion{N}{II}]/H$\alpha$, [\ion{O}{I}]/H$\alpha$, and [\ion{O}{III}]/H$\beta$ line ratios as a function of $\Sigma_{\rm H\alpha}$ in the DIG for our MW-like simulation in Fig.~\ref{fig:musecomparison}. The mock observations were split into five radial bins from from [0\,--\,16]\,kpc and the line ratio profile was calculated for each. In order to compare the line ratio trends between the radial bins and across snapshots, which have different average line ratios, each of the profiles is normalised such that the average line ratio in each radial bin is unity. The trendline represents the median values from four snapshots with five radial bins each.

We decided to show the trends in this way to facilitate a direct comparison with the PHANGS-MUSE observations, which are already split into five radial bins to isolate the effects of a metallicty gradient \citep{Belfiore:2022aa}. We compare our results to the PHANGS-MUSE galaxies which meet our MW-like criteria, which is as above; stellar masses of $\log(M_{\star}/\mathrm{M}_{\odot})=10.6$--$10.9$ and $\mathrm{SFR}/(\mathrm{M}_{\odot}\,\mathrm{yr}^{-1})=0.5$--$5.0$. This leaves us with a sample of six galaxies.

Our analysis reveals that the [\ion{S}{II}]/H$\alpha$, [\ion{O}{I}]/H$\alpha$, and [\ion{O}{III}]/H$\beta$ line ratio trends in our simulation agree well with those observed in the MW-like PHANGS-MUSE galaxies. Specifically, we capture the increase in these line ratios with decreasing surface brightness.

The [\ion{N}{II}]/H$\alpha$ is consistent with the observations within the scatter, however at face value the PHANGS-MUSE galaxies show a more marked decrease with increase surface brightness. Interestingly, the [\ion{N}{II}]/H$\alpha$ trend shows a significant amount of variety between different galaxies compared to the other emission line ratios, with some galaxies showing much flatter profiles or steeper profiles (see Fig. 11 in \citealt{Belfiore:2022aa}). Our simulation represents one realisation, and therefore it is not necessarily surprising that it may not recreate the [\ion{N}{II}]/H$\alpha$ profiles seen in the diverse range of MW-like galaxies. While we may not capture the full variety of [\ion{N}{II}]/H$\alpha$ behaviours, we can be confident that we do faithfully represent a subset of these systems. In summary, our simulation successfully reproduces key observational trends within the DIG, affirming its validity as a model for exploring the physical processes underlying these phenomena.

\section{Physics of the DIG}
\label{sec:PhysicsoftheDIG}

Having shown that we are able to reproduce key observational features of the DIG, we now seek to understand the physical properties of the ISM and radiative transfer effects that drive the observed line ratio trends.
We begin in Section~\ref{sec:PhysicalProperties} by discussing our approach to defining the DIG from a theoretical point of view.
In Section~\ref{sec:TheDIGinEmission}, we explore to what extent the observed line ratio trends are due to observational biases rather than reflecting the intrinsic emission of the DIG.
In Section~\ref{sec:ExplaininglineratiosintheDIG}, we investigate how changing emissivity and relative abundance of ionisation states leads to these trends.
In Section~\ref{sec:SourceofIonisingPhotons}, we discuss how \HII region leakage and HOLMES impact the distribution of ionisation states in the DIG. 
In Section~\ref{sec:Transporting radiation into the DIG}, we explore the impact of radiative transfer effects on the radiation field in the DIG.

\subsection{Physical properties}
\label{sec:PhysicalProperties}

As we have seen, the DIG is usually defined in high-resolution emission-line maps using morphological analysis. Other definitions have been used, particularly for lower resolution data, such as selecting spaxels based on the equivalent width of H$\alpha$. These definitions lack a robust theoretical foundation as the methods are rooted in phenomenology rather than the intrinsic properties/characteristics of the gas.

To address this gap, we adopt the electron density ($n_{\rm e}$) as a more theoretically motivated parameterisation because it nicely connects observational and theoretical interests in the DIG. Observationally, the DIG is defined by its low surface brightness emission, which is closely related to the electron column density. Theoretically, this definition is appealing because $n_{\rm e}$ is a multiplicative factor in the rates for important physical processes, such as collisional excitation, collisional ionisation, and recombination. While $n_{\rm e}$ closely reflects $n_\text{\HII}$, it also traces the presence of any ionized elements. Another key benefit of using $n_{\rm e}$ is that we exclude gas which is dense but is weakly ionised, which would be incorrectly characterised as part of an \HII region if we simply used a measure of gas density, such as $n_H$.

\begin{figure} 
\centering
	\includegraphics[width=\columnwidth]{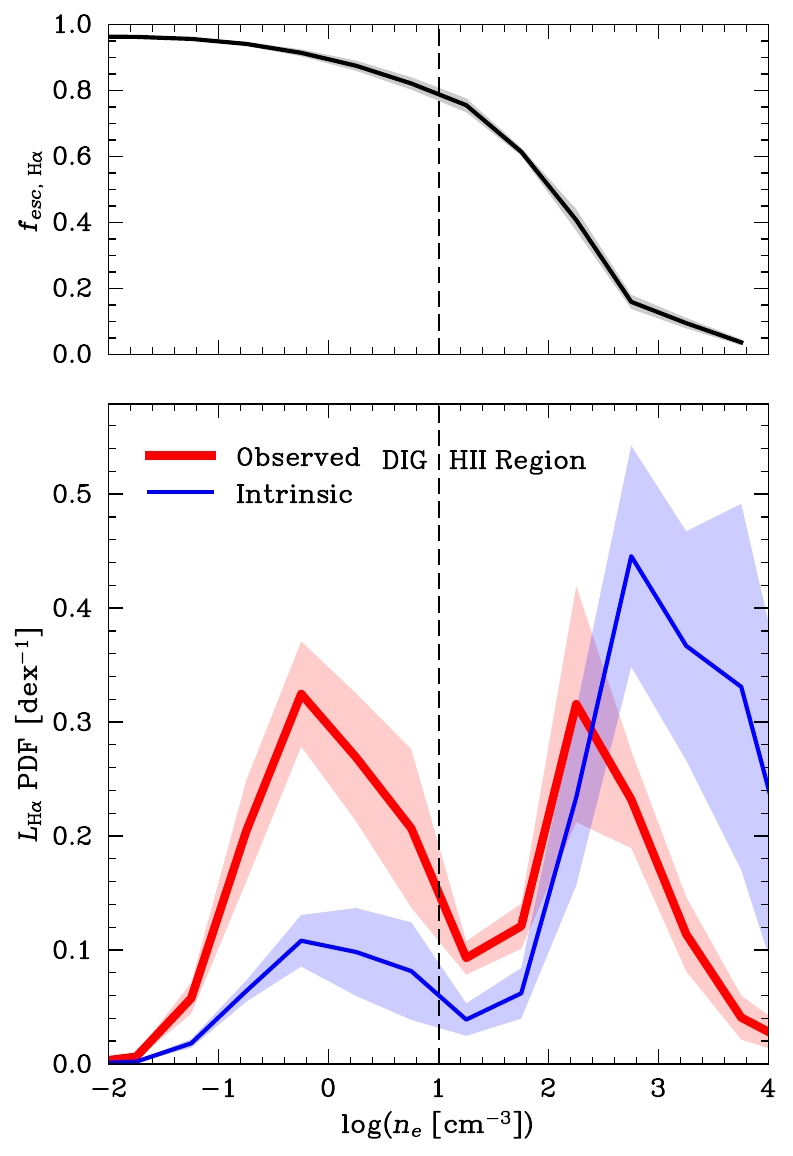}
    \caption{The emission and escape of H$\alpha$ emission in the DIG and \HII regions. The trend lines and shaded regions represent the mean and standard deviation across four snapshots. 
    \textit{Top:} The escape fraction of H$\alpha$ as a function of $n_{\rm e}$. We can see that \HII regions are intrinsically much brighter than the DIG, however the observed luminosities are comparable due to much higher dust attenuation in \HII regions causing low escape fractions.\textit{Bottom:} The distribution of intrinsic and observed H$\alpha$ luminosity of gas cells in the MW-like simulation as a function of $n_{\rm e}$.
    We can see bimodality in both the intrinsic and observed $L_{\mathrm{H_\alpha}}$, with two peaks at densities of $0.1$\,--\,$1\,\mathrm{cm}^{-3}$ and $10^2$\,--\,$10^3\,\mathrm{cm}^{-3}$. This motivates our use of 10 $\mathrm{cm}^{-3}$ as an electron density threshold between the DIG and \HII regions.}
    \label{fig:emission_ne}
\end{figure}

In Fig.~\ref{fig:emission_ne}, we plot the intrinsic and observed H$\alpha$ luminosity as a function of $n_{\rm e}$. There is a striking bimodality, with two peaks at $0.1$\,--\,$1\,\mathrm{cm}^{-3}$ and $10^2$\,--\,$10^3\,\mathrm{cm}^{-3}$ contributing the bulk of the observed H$\alpha$ emission. This is because \HII regions and the DIG are distinct physical phenomena, rather than two ends of a continuous $n_{\rm e}$ distribution. We can take advantage of this bimodality to robustly separate the DIG from \HII regions with an $n_{\rm e}$ threshold of $n_{\rm e,thresh}=10\,\mathrm{cm}^{-3}$. We delay our discussion of how these components arise until Section~\ref{sec:SourceofIonisingPhotons}, where we discuss the sources of ionisation, and instead now focus on the emission and properties of the DIG.

Despite contributing only 26 per cent of the intrinsic H$\alpha$ emission, the DIG accounts for 60 per cent of the observed emission. The primary reason for this is that H$\alpha$ emitted from \HII regions is much more prone to dust attenuation; $\sim80$ per cent of H$\alpha$ from \HII regions is absorbed by dust compared to $\sim10$ per cent of H$\alpha$ from the DIG. This is in agreement with observations of DIG H$\alpha$ emission, as discussed in \citet{Tacchella:2022aa}.

In the right panel of Fig.~\ref{fig:dendrogram}, we show the fraction of observed H$\alpha$ emission from cells which we define as \HII regions ($n_{\rm e}>10\,\mathrm{cm}^{-3}$) in each spaxel. We have overlaid the dendrogram, which was selected using the observed mock MUSE H$\alpha$ emission-line map. While it is encouraging that the our observational techniques select most of the spaxels dominated by \HII emission, it is also evident that there are relatively few of these spaxels. This makes sense when considering that the DIG accounts for 60\% of observed H$\alpha$, so unless there are many bright \HII regions along a line of sight, the amount of DIG contamination in a spaxel will be significant.

We can see in Fig.~\ref{fig:mw_properties} that recent star formation tends to be very clumpy. Regions of intense star formation, such as in the center of the galaxy, correspond to the brightest \HII regions identified. However, these regions also host bright clumps of DIG, which are easily misidentified by the dendrogram. We can also see much less intense clumps of star formation in Fig.~\ref{fig:mw_properties}, and these dimmer \HII regions are likely to be outshone or contaminated by the DIG along the line of sight. While DIG contamination of \HII region observations is a well-known effect, the scarcity of pure-\HII region spaxels, and how even regions which appear to be bona fide \HII regions can in fact be almost entirely due to DIG emission, should be cause for concern. We discuss this further in Section~\ref{sec:Discussion}.

\subsection{Line ratios in the DIG}
\label{sec:TheDIGinEmission}

The immediate question to answer is whether the trends in the line ratios shown in Fig.~\ref{fig:musecomparison} correspond to differences in line ratios between the DIG and \ion{H}{II} regions, and to what extent the trends are merely artefacts of observational biases, such as projection effects, or due to physical processes, such as dust scattering.

\begin{figure} 
\centering
	\includegraphics[width=\columnwidth]{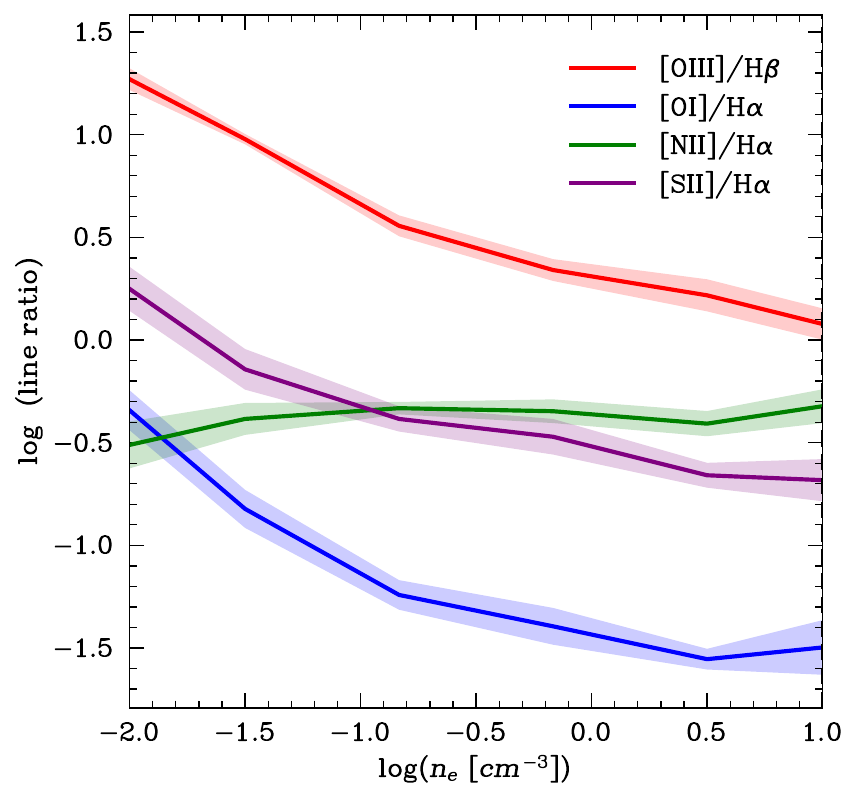}
    \caption{The [\ion{S}{II}]/H$\alpha$, [\ion{N}{II}]/H$\alpha$, [\ion{O}{I}]/H$\alpha$, and [\ion{O}{III}]/H$\beta$ line ratios as a function of $n_{\rm e}$. The trend line and shaded region represent the mean and standard deviation (in log space) across four snapshots. Line ratios were calculated by summing the observed luminosity in a metal emission line of all gas cells in the density bin, and then dividing by the summed observed luminosity of H$\alpha$ or H$\beta$ as appropriate. We can clearly see the same line ratio trends as in Fig.~\ref{fig:musecomparison}, however with a much steeper slope.}
    \label{fig:emissionratios_ne}
\end{figure}

To answer this, we analyze the gas by binning the cells according to $n_{\rm e}$ and calculating the observed luminosity for each bin. These luminosities enable us to calculate line ratios for the gas at each density. This is essentially the theoretical analogue of the surface brightness plots in Fig.~\ref{fig:musecomparison}.

In Fig.~\ref{fig:emissionratios_ne}, we show the line ratios against $n_{\rm e}$ and confirm that the observed trends in these line ratios indeed correlate with changes in ionised gas density, particularly with decreasing $n_{\rm e}$. We have not plotted the intrinsic line ratios as they are nearly indistinguishable from the observed line ratios since dust attenuation has little impact on the ratio of two lines with similar wavelengths. 

While the trends are the same, it is interesting to see that the line ratios reach much more extreme values than is seen in our mock observations. This arises from the integration of photons emitted at different gas densities along the observed line of sight, compounded by dust scattering that further mixes photons from different gas densities.

The observed surface brightness is a luminosity-weighted average of the ionised gas along the line of sight. Even though mixing causes the observed emission-line ratios to be less extreme, the general trends are preserved because lower surface brightnesses correlate with lower luminosity-weighted average densities along the line of sight.

\subsection{Explaining line ratios in the DIG}
\label{sec:ExplaininglineratiosintheDIG}

Having shown that [\ion{S}{II}]/H$\alpha$, [\ion{O}{I}]/H$\alpha$, and [\ion{O}{III}]/H$\beta$ intrinsically increase as the ionised gas becomes more diffuse, rather than being driven by observational biases, we can now explore how these trends arise.

Line ratios are fundamentally dependent on two factors; the relative emissivity of the emission lines and the relative abundance of the ions which give rise to the line ratios. As discussed in Section~\ref{sec:Introduction}, both of these factors have been considered but it has been difficult to disentangle the degeneracy between them through observations. Fortunately, with our post-processing approach, we are able to disentangle the impacts of emissivity and abundance.

\begin{figure} 
\centering
	\includegraphics[width=\columnwidth]{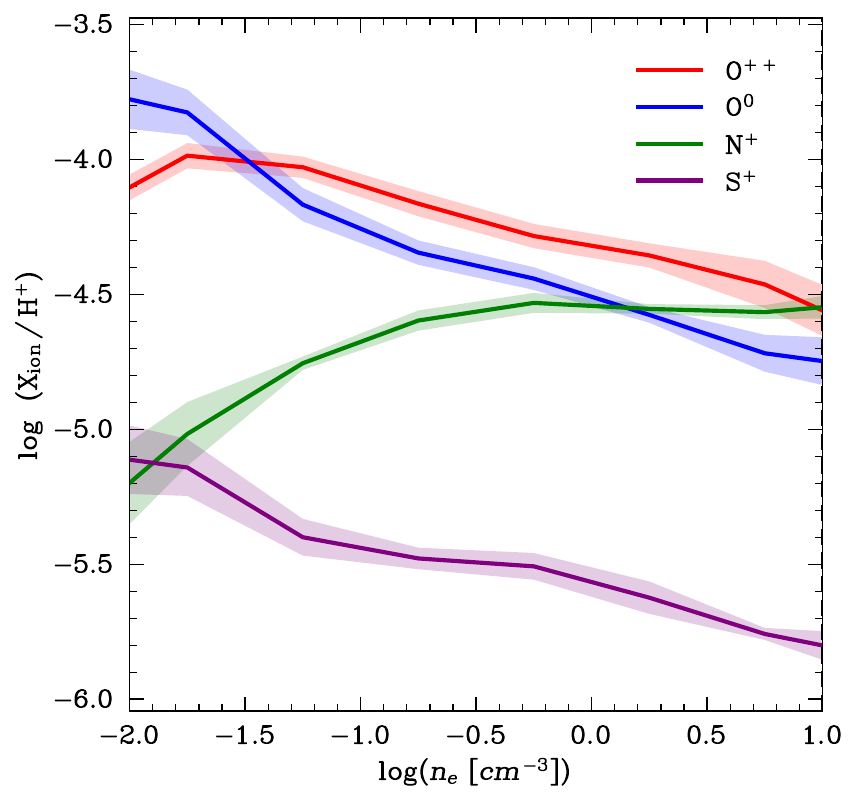}
    \caption{The abundance of $\mathrm{S^{+}}$, $\mathrm{O^{0}}$, $\mathrm{N^{+}}$, and $\mathrm{O^{++}}$ relative to $\mathrm{H^{+}}$ is shown as a function of $n_{\rm e}$. The trend line and shaded region represent the mean and standard deviation across four snapshots. $\mathrm{S^{+}}$/$\mathrm{H^{+}}$, $\mathrm{O^{0}}$/$\mathrm{H^{+}}$, and $\mathrm{O^{++}}$/$\mathrm{H^{+}}$ all increase with decreasing $n_{\rm e}$, consistent with the trends of their line ratios, [\ion{S}{II}]/H$\alpha$, [\ion{O}{I}]/H$\alpha$, and [\ion{O}{III}]/H$\beta$, although with different magnitudes and curvature. $\mathrm{N^{+}}$/$\mathrm{H^{+}}$ decreases with decreasing $n_{\rm e}$, in contrast to the flat [\ion{N}{II}]/H$\alpha$ trend. This shows the importance of jointly considering emissivity (Fig.~\ref{fig:emissivity_ne}) and abundances (shown here).}
    \label{fig:abundance_ne}
\end{figure}

We will first consider the impact of relative abundance, and so in Fig.~\ref{fig:abundance_ne} we show the abundance of $\mathrm{S^{+}}$, $\mathrm{O^{0}}$, $\mathrm{N^{+}}$, and $\mathrm{O^{++}}$ relative to $\mathrm{H^{+}}$ as a function of $n_{\rm e}$. Interestingly, we find that the $\mathrm{O^{0}}$/$\mathrm{H^{+}}$, $\mathrm{S^{+}}$/$\mathrm{H^{+}}$, and $\mathrm{O^{++}}$/$\mathrm{H^{+}}$ ratios all increase with decreasing $n_{\rm e}$, which matches the trends of their respective line ratios. However, the shape of these curves does not match exactly with the line ratio trend. Conversely, the $\mathrm{N^{+}}$/$\mathrm{H^{+}}$ ratio decreases with decreasing $n_{\rm e}$, whereas the trend of [\ion{N}{II}]/H$\alpha$ is flat. This indicates that the abundance ratios alone do not tell the full story and that emissivity must also be considered.

In discussing these line ratio trends, it is important to note that collisional ionisation and the UV background have a negligible impact. Charge exchange is dominant for the ionisation of $\mathrm{O^{0}}$, subdominant for $\mathrm{N^{+}}$, and unimportant for $\mathrm{S}$ and $\mathrm{O^{++}}$. This means that photoionisation is the primary ionisation source for $\mathrm{N^{0}}$, $\mathrm{S^{0}}$, and $\mathrm{O^{++}}$. The rising $\mathrm{O^{++}}$/$\mathrm{H^{+}}$ with decreasing $n_{\rm e}$ is a clear indicator of the hardening radiation field with decreasing $n_{\rm e}$, and we will investigate the causes of this in Section~\ref{sec:SourceofIonisingPhotons}.

\begin{figure} 
\centering
	\includegraphics[width=\columnwidth]{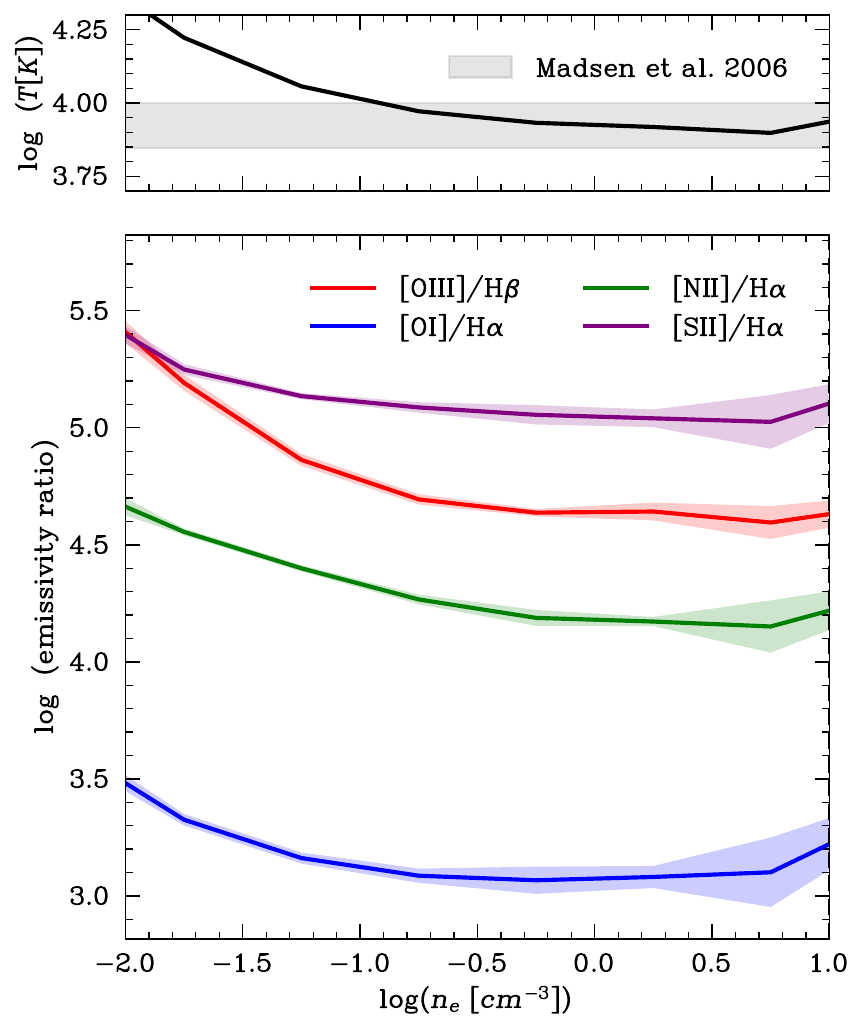}
    \caption{A comparison of the effective relative emissivity of the [\ion{S}{II}]/H$\alpha$, [\ion{N}{II}]/H$\alpha$, [\ion{O}{I}]/H$\alpha$, and [\ion{O}{III}]/H$\beta$ line ratios. The trend line and shaded region represent the mean and standard deviation across four snapshots (in log space). \textit{Top:} The average temperature of gas weighted by observed H$\alpha$ flux. The grey band shows the range of DIG temperatures observed in the Milky Way by \citet{Madsen:2006aa}. This shows that the temperature of the observable DIG is consistent with the observed DIG temperature. \textit{Bottom:} The relative emissivity of the line ratios generally increases with decreasing $n_{\rm e}$ due to the collisionally excited metal lines being more sensitive to the increasing temperature in the DIG. The changing emissivity is vital to drive the line ratio trends in Fig.~\ref{fig:emissionratios_ne}, together with the abundances shown in Fig.~\ref{fig:abundance_ne}.
    }
    \label{fig:emissivity_ne}
\end{figure}

In Fig.~\ref{fig:emissivity_ne} we show the change in relative emissivity for the line ratios. The emissivity ratios considered increase with decreasing $n_{\rm e}$ because of increasing temperature in the DIG with decreasing $n_{\rm e}$. The temperature increase causes a more rapid increase in emissivity for CELs than for RLs such as H$\alpha$ and H$\beta$. 

The temperatures of our DIG are consistent with the observed DIG in the Milky Way. \citet{Madsen:2006aa} find that the DIG within the Milky Way is in the range of [7000\,--\,10\,000]\,K, with an average of 9000K. We find the average temperature of our DIG is 9000\,K when weighted by observed H$\alpha$ emission to account for observability. Fig.~\ref{fig:emissivity_ne} also shows that the vast majority of luminous gas (see Fig.~\ref{fig:emission_ne}) is within observed Milky Way DIG temperature ranges. The important insight here is that the trend of increasing temperature with $n_{\rm e}$ is enough to boost the relative emissivity of the line ratios with decreasing $n_{\rm e}$.

\subsection{Source of ionising photons}
\label{sec:SourceofIonisingPhotons}

We now aim to understand the hardening of the radiation field with decreasing $n_{\rm e}$, which is implied by the changes in ionic abundances presented in Fig.~\ref{fig:abundance_ne}. We focus on the hardening of radiation in the DIG in reference to the fraction of photons with energy above the ionisation energy of $\mathrm{O^{++}}$, 35.12\,eV. Quantities regarding hard ionising photons with energy above 35.12\,eV are denoted with subscript $\mathrm{h}$, whereas rates regarding all ionising photons with energy above 13.6\,eV are denoted with subscript $0$.

\begin{figure}
\centering
	\includegraphics[width=\columnwidth]{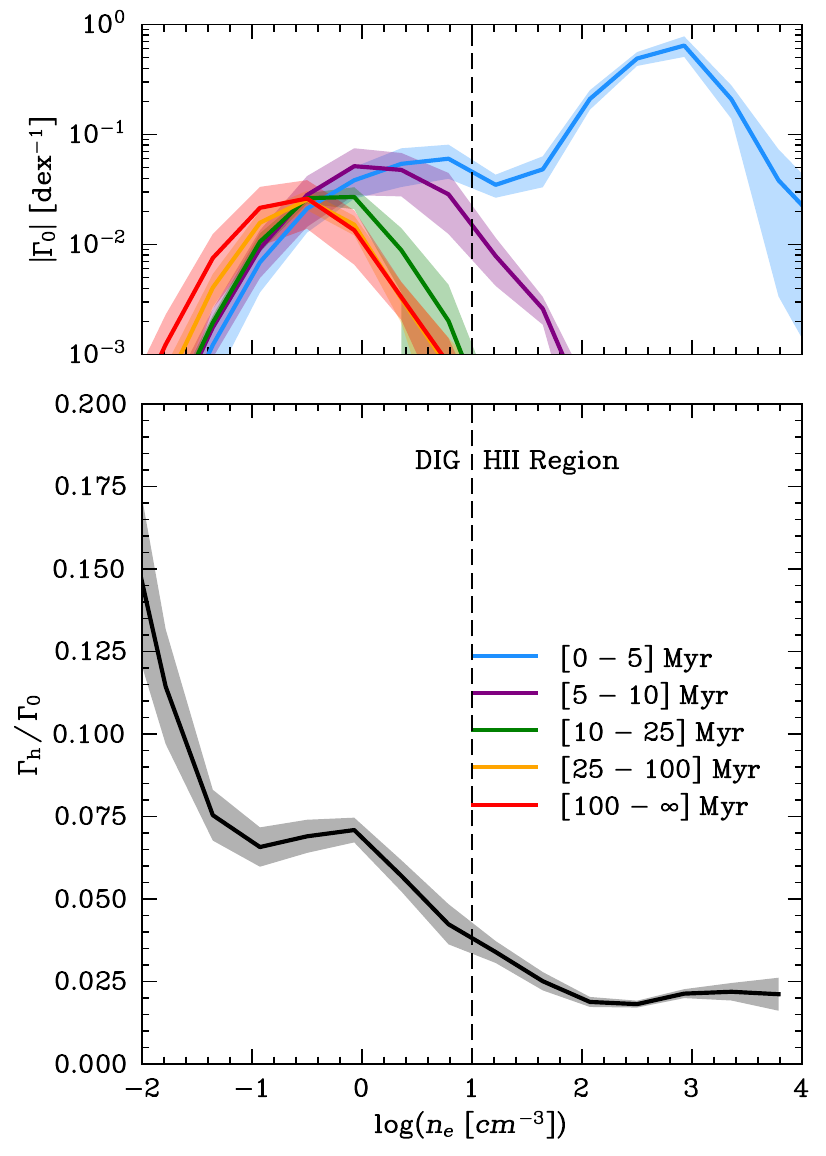}
    \caption{The evolution of the ionising radiation field with $n_{\rm e}$. The trend line and shaded region represent the mean and standard deviation across four snapshots. \textit{Top:} The rate of photoionisations due to photons above the \HI ionisation energy (13.6\,eV) for a given age bin, $\Gamma_0$, normalised by the total number in the simulation, $\sum \Gamma_0$. Stars aged younger than 5\,Myr are the dominant source of ionisation in \HII regions. As $n_{\rm e}$ decreases, there is a transition of the dominant ionising source to older stars.
    \textit{Bottom:} The rate of photoionisations due to photons above the \OII ionisation energy (35.12\,eV), $\Gamma_\mathrm{h}$, divided by $\Gamma_0$. The ionising radiation field becomes harder with decreasing $n_{\rm e}$ because the dominant source of ionising radiation transitions to older stars which have intrinsically harder SEDs.}
    \label{fig:radiation_hardness_overall}
\end{figure}

The evolution of the radiation field from \HII regions to the DIG is presented in the bottom panel of Fig.~\ref{fig:radiation_hardness_overall}, where we show the fraction of photoionisations due to hard ionising photons, $\Gamma_\mathrm{h}/\Gamma_0$, as a function of $n_{\rm e}$. This is essentially a measure of the effective hardness of the ionising radiation field at particular electron density. The overall radiation field in the diffuse gas becomes harder, with $\Gamma_\mathrm{h}/\Gamma_0$ increasing from 0.03 at $n_{\rm e}=10\,\mathrm{cm^{-3}}$ to 0.14 at $n_{\rm e}=10^{-2}\,\mathrm{cm^{-3}}$. The top panel of Fig.~\ref{fig:radiation_hardness_overall} shows the normalised rate of photoionisations above the \HI ionisation energy, $|\Gamma_0|$, as a function of $n_{\rm e}$.

\begin{figure}
\centering
	\includegraphics[width=\columnwidth]{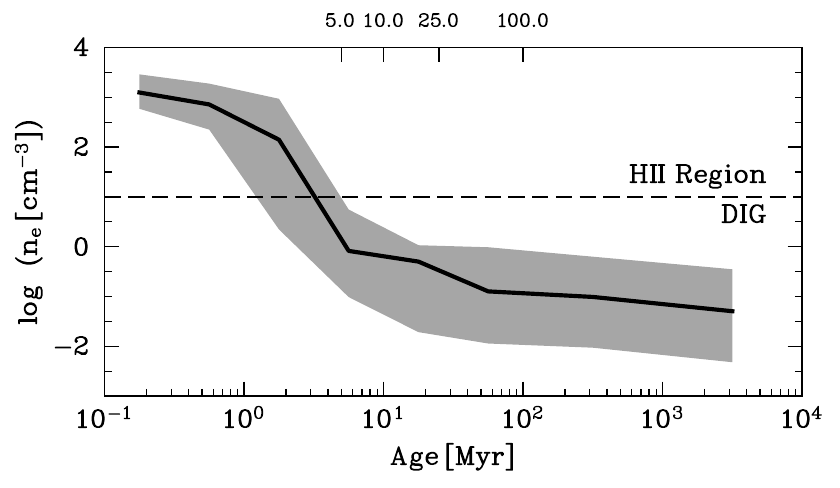}
    \caption{The median $n_{\rm e}$ of gas immediately surrounding stars of different ages. Shaded regions represent the 16-84$\rm ^{th}$ percentile scatter. Stars have cleared their dense \HII regions by $\sim5$\,Myr age through radiation pressure and nearby supernovae. This exposes them to the DIG while they are still emitting a copious amount of ionising radiation, allowing stars of ages between 5 and 25\,Myr to ionise the DIG.}
    \label{fig:star_density}
\end{figure}

\begin{figure}
\centering
	\includegraphics[width=\columnwidth]{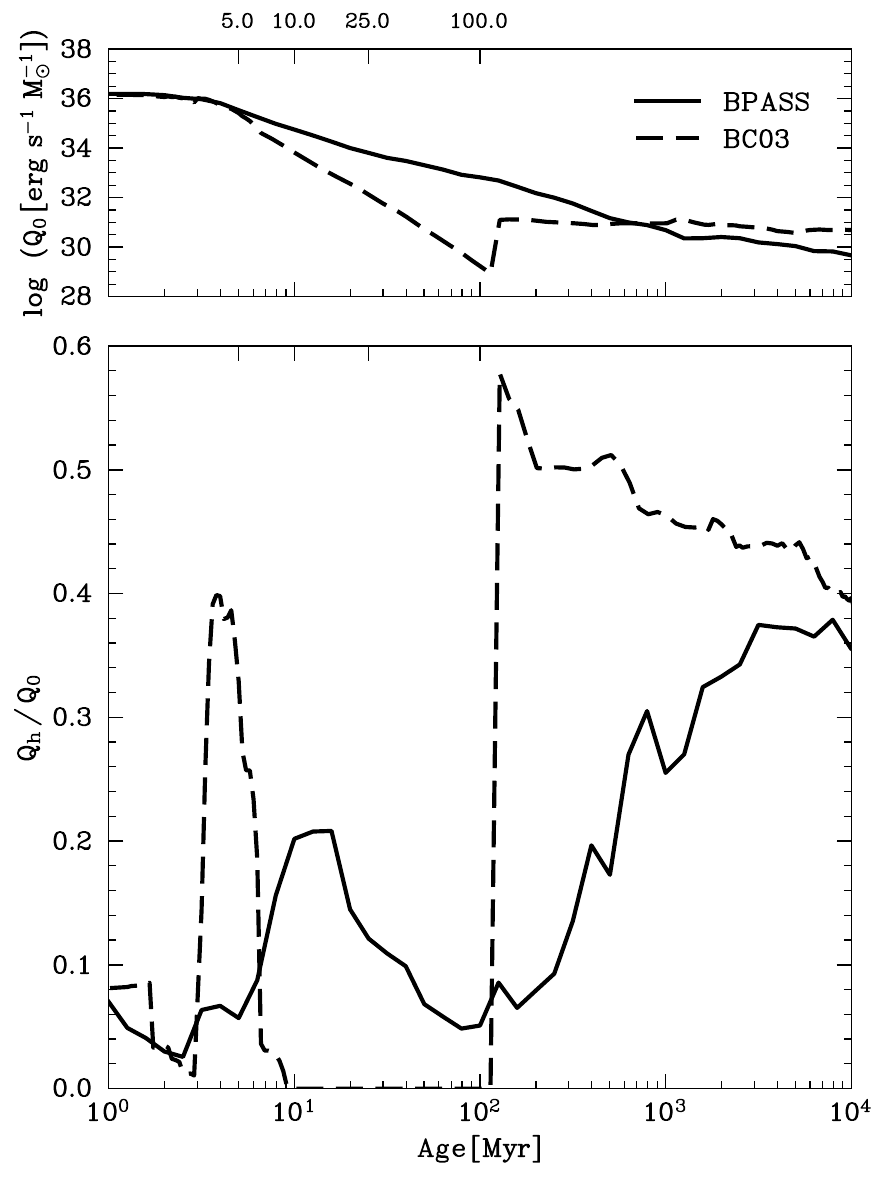}
    \caption{The intrinsic properties of stars in the simulation. We use the BPASS library (solid lines), however we plot the BC03 properties (dashed lines) as a comparison. \textit{Top:} The ionising emissivity of stars as a function of age. The youngest stars dominate the production of ionising photons. \textit{Bottom:} The rate of O$^{++}$ ionising photons (>35.12\,eV) from stars, $\mathrm{Q_\mathrm{h}}$, divided by the rate of \HI ionising photons (>13.6\,eV) from stars, $\mathrm{Q_0}$. The hardness of the ionising radiation from stars in the BPASS library peaks from 5 to 25 Myr (a shorter and earlier hardness peak is also seen in BC03). This peak explains importance of these stars in driving a hard radiation field in the DIG, when considered alongside the low-density environments in which these stars reside (Fig.~\ref{fig:star_density}).}
    \label{fig:bpass}
\end{figure}

Two primary mechanisms could account for this hardening of the radiation field with decreasing $n_{\rm e}$: either the ionising radiation emitted from sources in the dense gas (e.g. young stars in \HII regions) becomes harder as it travels into the DIG, or the intrinsic hardness of the ionising sources change because of a relative change in the contribution from stars of different ages. 

Due to our Monte Carlo approach, we can trace the source of each photon absorbed in the gas cells. This allows us to dissect the contribution of various stellar populations to gas ionisation as a function of density, and therefore isolate the impact of stars of different ages. We have selected several age bins which define different stellar populations of interest, [0\,--\,5]\,Myr, [5\,--\,10]\,Myr, [10\,--\,25]\,Myr, [25\,--\,100]\,Myr, and [100\,--\,$\infty$]\,Myr. 

The [0\,--\,5]\,Myr population represents the youngest stars, recently born and still residing in dense gas, as shown in Fig.~\ref{fig:star_density}. In the top panel of Fig.~\ref{fig:bpass}, we show that these stars emit copious amounts of \HI ionising photons, contributing 84\% of the overall ionising budget. The bottom panel shows that the intrinsic spectrum of these stars is soft. Looking at the top panel of Fig.~\ref{fig:radiation_hardness_overall}, we can see that the [0\,--\,5]\,Myr population is nearly entirely responsible for ionising \HII regions. This is unsurprising given that this is the only population which resides in high-density ionised gas and it has the highest intrinsic ionising luminosity. Turning to the bottom panel, the overall hardness of radiation causing photoionisation in \HII regions exactly tracks the contribution from these stars. The radiation here is directly injected at \HII densities by these young stars and therefore reflects their soft intrinsic spectrum. Interestingly, despite being so intrinsically luminous these stars contribute only a modest 26\% of the ionising photons absorbed in the DIG. This implies relatively little leakage from the \HII regions surrounding these stars, with only 18\% escaping to be absorbed in the DIG.

The [5\,--\,10]\,Myr population also emits a copious amount of ionising radiation, albeit less so than their younger counterparts, as shown in top panel of Fig.~\ref{fig:bpass}. In the bottom panel, we can also see that the intrinsic hardness of their emitted radiation significantly increases. In Fig.~\ref{fig:star_density}, we see that the dense \HII regions have mostly dissipated by the time stars become 5\,Myr old, through a combination of early stellar and local supernovae feedback. Despite being less intrinsically luminous than the younger stars in ionising photons, this population contributes even more strongly to DIG ionisation (29\%) because the radiation is less shielded by high-density gas. Additionally, their intrinsically harder ionising SED causes these stars to contribute very strongly to the hard ionising radiation in the DIG (32\%).

Stars aged [10\,--\,25]\,Myr are situated firmly within the DIG (Fig.~\ref{fig:star_density}), however because of their lower intrinsic luminosity these stars contribute less to the ionisation of the DIG. This population is vital in understanding the properties of the DIG because it emits hard ionising radiation, as shown in Fig.~\ref{fig:bpass}. This means that these stars drive much of the hardening in the DIG, contributing 38\% of the hard ionising photons absorbed in the DIG compared to 21\% of the total \HI ionising photons. The hardened radiation of these stars is due to the Wolf-Rayet phase, which we discuss further in Section~\ref{sec:Discussion}.

The ionising luminosity and hardness falls off rapidly after 25\,Myr, such that the [25\,--\,100]\,Myr population is not particularly important to the DIG radiation field in this simulation, emitting a small amount of relatively soft ionising radiation into the DIG.

The [100\,--\,$\infty$]\,Myr population includes many different types of stars, but most notably the HOLMES or post-AGB stars which are thought to release a significant amount of hard ionising photons. Although the age range of this population is broad and includes stars which may not usually be thought of as ``old'', it is appropriate for our purposes because we are still separating the HOLMES from other potentially important populations. The oldest stars in this population emit the hardest radiation, particularly at ages greater that 1\,Gyr, as seen in the bottom panel of Fig.~\ref{fig:bpass}. However, we find that these stars are only a very minor contribution to the ionising radiation field in the DIG, primarily because of the very low intrinsic ionising luminosity.

When considering the roles of these different populations, we can understand how the radiation field hardens as the gas becomes more diffuse. Very young stars (<5\,Myr) are nearly entirely responsible for the ionisation of \HII regions, however radiation leaking from these \HII regions contributes modestly to the overall ionisation of the DIG (26\%) and is a small contribution to the hard radiation absorbed in the DIG (12\%). After $\sim5$\,Myr, the \HII regions are dispersed by feedback from the massive stars and nearby supernovae, meaning that this population is situated in the DIG. The clearing of \HII regions coincides with an uptick in hardness of the intrinsic SED, allowing these still young and luminous stars to propagate a hard radiation field into the DIG.

As the stars age, their ionising SED hardens and they emit less ionising radiation. We can see in the top panel of Fig.~\ref{fig:radiation_hardness_overall} that the [0\,--\,5]\,Myr, [5\,--\,10]\,Myr, and [10\,--\,25]\,Myr populations are dominant in increasingly diffuse gas as their star formation sites become disrupted by stellar feedback and they are more firmly situated in diffuse gas. As the intrinsic hardness of the radiation also increases from [0\,--\,5]\,Myr, [5\,--\,10]\,Myr, to [10\,--\,25]\,Myr, this causes the a hardening of the radiation field. The ionising output of the stars falls off rapidly such that all stars above 25\,Myr contribute only relatively modestly to DIG ionisation, and are not functionally important for ionising the DIG.

The relative balance of these populations would change the picture in terms of contribution of stars at different ages to the ionisation of the DIG. For example, if our simulation had many more stars in the [5\,--\,10]\,Myr age range and fewer in the [10\,--\,25]\,Myr range, we would expect the radiation to harden less with decreasing $n_{\rm e}$. However, we would expect that the overall trend of an increasingly hard radiation field in the DIG, driven by recent star formation, remains the same. It is interesting that the PHANGS-MUSE observations show significant scatter in the line ratio trends between different galaxies \citep{Belfiore:2022aa}. It is possible that much of this scatter could be driven by the precise star-formation history within the past 25\,Myr.

\subsection{Transporting radiation into the DIG}
\label{sec:Transporting radiation into the DIG}

While observations show great variety in DIG properties between different galaxies, there is remarkable consistency radially within each galaxy when correcting for a metallicity gradient \citep{Belfiore:2022aa}. This could conflict with the explanation that the DIG properties are driven by recent star formation because we would expect there to be a large variation in the local star-formation history from region to region within a galaxy, which may show up as scatter in the DIG properties in different regions or radial bins of the galaxy.

\begin{figure}
\centering
	\includegraphics[width=\columnwidth]{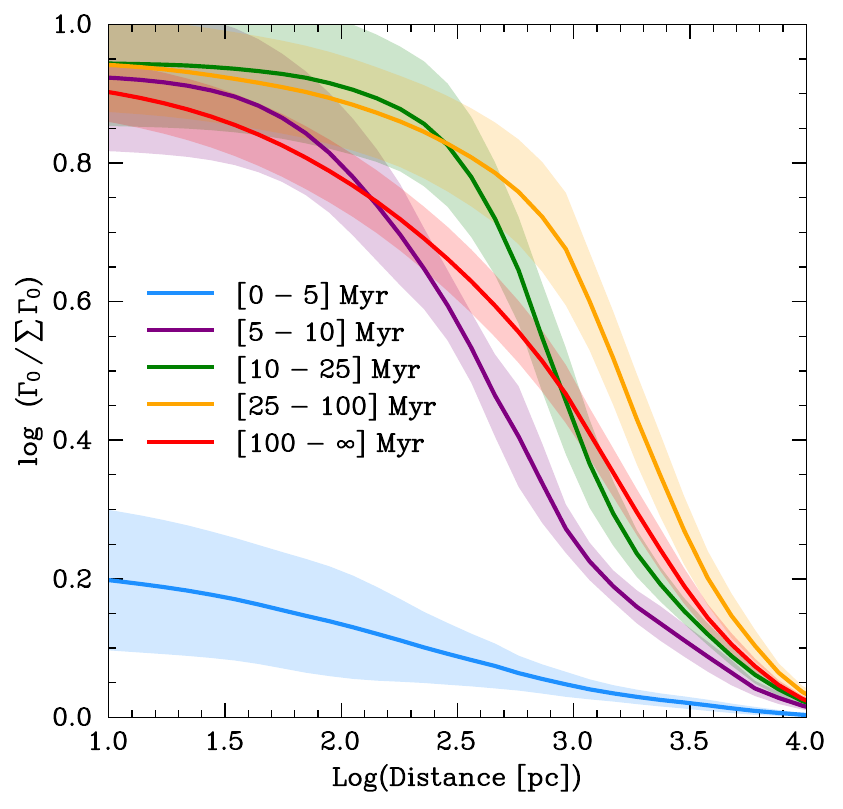}
    \caption{The fraction of ionising photons which have not yet been absorbed as function of distance from the source star. The trend line and shaded region represent the mean and standard deviation across four snapshots. Stars younger than 5\,Myr emit ionising photons which are mostly absorbed on small scales (<1\,pc), however photons which leak from \HII regions can propagate on kpc scales. More importantly, ionising photons from stars older than 5\,Myr propagate across scales up to 10\,kpc. This long mean free path allows stars of ages between 5 to 25\,Myr to ionise the DIG on galactic scales.}
    \label{fig:age_dist_freq}
\end{figure}

However, DIG properties can be driven by recent star formation and be consistent throughout a galaxy if the DIG properties are set over many kiloparsec. This is because, on these scales, the local variations in star-formation history can be smoothed out (see Fig.~\ref{fig:mw_properties}). We investigate this in Fig.~\ref{fig:age_dist_freq}, where we show the fraction of ionising photons which have not yet been absorbed as function of distance from the source star.

For stars still embedded in \HII regions (<5\,Myr) most radiation is absorbed on scales comparable to \HII region sizes (10\,pc). However, when radiation does leak from \HII regions, it  propagates to large distances. Ionising photons emitted from stars 5\,Myr old are absorbed on even larger scales, and in particular, 33\% and 40\% of the ionising photons from the [5\,--\,10]\,Myr and [10\,--\,25]\,Myr populations are absorbed on distances of 1\,--\,10\,kpc. Given that these populations drive the bulk of DIG ionisation, we conclude that recent star formation can drive uniform DIG properties on galaxy-wide scales.
 
\section{Discussion}
\label{sec:Discussion}

It is challenging to isolate the emission from \HII regions and the DIG in observations. Through the use of techniques such as dendrogram selection, observers select peaks in H$\alpha$ surface brightness (applied to our simulation in Fig.~\ref{fig:dendrogram}). While the technique does well at identifying most \HII dominated pixels, it also misidentifies many regions which are in fact DIG dominated. The fraction of false identifications can be improved by adjusting dendrogram parameters, generally by masking lower surface brightness pixels. However, even if it were possible to achieve 100$\%$ accuracy in identifying \HII regions, the fundamental problem is that there are few pixels which are truly \HII dominated due to the projection of the DIG and \HII regions along the line of sight (see Fig.~\ref{fig:dendrogram}). Fainter \HII region are unable to outshine the DIG, and therefore we should be careful when drawing trends with the brightness of \HII regions. At the least, we caution against pushing to low H$\alpha$ surface brightness when identifying \HII regions on high resolution emission line maps and we strongly advocate that emission from the DIG is considered when analysing the line ratios from extragalactic \HII regions.

Despite the challenges of isolating the emission from \HII regions and the DIG in observations, we find that the observed line ratio trends with H$\alpha$ surface brightness in the DIG match the intrinsic trends of these line ratios with $n_\mathrm{e}$. These trends have mostly been attributed to changing relative abundances of ionisation states in the DIG, which we find is an important factor. However, since more diffuse gas is generally hotter, changes in relative emissivity between collisional metal lines and the hydrogen recombination lines with temperature also contribute to the observed trends. This is a particularly important aspect to consider when carrying out 1D photoionisation modelling, where it may be difficult to set the temperature in the DIG in a self-consistent way.

The line ratio trends, in particular the increase in [\ion{O}{III}]/H$\beta$ with decreasing $n_\mathrm{e}$, implies hardening ionising radiation in the DIG. The favoured explanation for this has been the presence of HOLMES in the DIG which comprise a small fraction of the overall ionisation, but emit an outsize amount of high energy photons \citep{Zhang:2017aa, Belfiore:2022aa}. We instead find that the ionisation of the DIG is driven by stars of ages between 5 and 25\,Myr, which have generally cleared their \HII regions and emit hard ionising radiation.

The hardened radiation in the [5\,--\,25]\,Myr age range is caused by Wolf-Rayet (WR) phase stars. WR stars are hotter than O type stars, and thus emit harder ionising radiation. Importantly, the BPASS stellar models used in this study account for the creation of WR stars through binary interactions which strip the hydrogen envelope of cool red supergiants, which creates many more WR stars at older ages (up to $\sim$100\,Myr) compared to single-star modelling \citep{Xiao:2018aa}.

Using the single-star \citet{Bruzual:2003aa} (BC03) model instead would lead to $\sim$2 orders of magnitude less intrinsic ionising luminosity for the [10\,--\,100]\,Myr stars. Interestingly, Fig.~\ref{fig:bpass} shows that that the BC03 stellar population synthesis model also shows an even stronger uptick in the hardness of stellar ionising spectra, however only between 5 to 10\,Myr. In principle, these stars being even more luminous and with even intrinsically harder ionising radiation may be sufficient to ionise the DIG with a hard radiation field. However, we have not investigated whether this would be sufficient to reproduce the DIG line ratio trends. Additionally, this would place even greater importance on the early clearing of \HII regions in order to allow for a significant amount of leakage on $\sim$5\,Myr timescales. Our results indicate that this could be the case (Fig.~\ref{fig:star_density}), in agreement with previous observational studies \citep{Kim:2021aa, Hannon:2022aa, Chevance:2022aa}.

The dependence of DIG properties, such as line ratios, on the ionising SEDs of stars at all ages and on the escape of ionising radiation from young stars in particular provides an exciting opportunity to constrain stellar feedback and stellar population models. For example, in addition to binary stars, rotating stars can also prolong the emission of ionising photons \citep{Choi:2017aa}. However, while there are differences between the intrinsic SEDs of binary and rotating stars, they can both generally reproduce the observed line ratios in \HII regions \citep{Byler:2017aa,Xiao:2018aa}. Due to the increased sensitivity to the SEDs of stars beyond 5\,Myr, DIG line ratios may provide a more potent tool to differentiate the models.

While our study has focused on galaxies in the local universe, it may have interesting implications for the Epoch of Reionisation. At these redshifts ($z>5$), we may expect diffuse gas to be less prevalent. If our results hold, with much of the ionising photons beyond 5\,Myr escaping \HII regions, then we would expect a significant fraction of the ionising photons to escape the ISM. This may provide an explanation for observed ``mini-quenched'' galaxies, which show signs of recent star formation but unexpectedly little line emission \citep{Looser:2023ab,Dome:2024aa}. However, our results depend on the stellar SEDs and physical processes in the ISM such as cooling, which do vary with metallicity, and thus it is difficult to extrapolate our results to the high redshift regime.

\section{Conclusions}
\label{sec:Conclusions}

We have undertaken a pioneering self-consistent galaxy-scale study of the ionisation and emission from the diffuse ionised gas in star-forming galaxies. This level of detailed modelling is necessary to capture the interplay of different stellar populations, the complex radiative transfer effects, and the varied heating sources present throughout the ISM.
 
Our analysis is based on a high-resolution simulation of a Milky Way-like galaxy, which includes state-of-the-art physics such as multi-band on-the-fly radiative transfer, dust formation and destruction, and a realistic stellar feedback subgrid recipe. We employed the Monte Carlo radiative transfer (MCRT) code \textsc{colt} to post-process the radiative transfer of ionising and emission line photons, and we carefully compared our mock observations to MaNGA and to $\sim$10\,pc resolution MUSE emission line maps to confirm that we produce a physically realistically DIG in emission (Fig.~\ref{fig:musecomparison}).

Our main conclusions are summarised as follows:
\begin{enumerate}
\item The DIG and \HII regions are distinct when the observed $\mathrm{L_{H\alpha}}$ from gas cells is plotted against electron density, $n_{\rm e}$, as we can see that the bulk of observed H$\alpha$ emission is due to two peaks at $0.1$\,--\,$1\,\mathrm{cm}^{-3}$ and $10^2$\,--\,$10^3\,\mathrm{cm}^{-3}$ (Fig.~\ref{fig:emission_ne}). We use this bimodality to define the DIG/\HII region threshold as $n_{\rm e}=10\,\mathrm{cm^{-3}}$.
\item The observed trends in [\ion{S}{II}]/H$\alpha$, [\ion{N}{II}]/H$\alpha$, [\ion{O}{I}]/H$\alpha$, and [\ion{O}{III}]/H$\beta$ with decreasing surface brightness correspond to genuine trends in these emission line ratios as ionised gas becomes more diffuse; i.e. with decreasing $n_{\rm e}$ (Fig.~\ref{fig:emissionratios_ne}). These trends are caused by a combination of changing relative emissivity and relative abundances with decreasing $n_{\rm e}$.
\item The relative emissivity of the [\ion{S}{II}]/H$\alpha$, [\ion{N}{II}]/H$\alpha$, [\ion{O}{I}]/H$\alpha$, and [\ion{O}{III}]/H$\beta$ line ratios increase with decreasing $n_{\rm e}$ due to an increasing temperature with decreasing $n_{\rm e}$ (Fig.~\ref{fig:emissivity_ne}). The increase in $\mathrm{O^{++}}$ relative to $\mathrm{H^{+}}$ with decreasing $n_{\rm e}$ is caused by the radiation field becoming harder with decreasing $n_{\rm e}$. 

\item The DIG is primarily ionised by radiation from stars recently formed stars of ages between 5 and 25\,Myr. Ionisation of the DIG by these stars is primarily possible due to clearing of \HII regions by $\sim5$\,Myr (Fig.~\ref{fig:star_density}), exposing bright young stars directly to the DIG, rather than bona fide leakage through dense \HII regions (Fig.~\ref{fig:radiation_hardness_overall}).

\item As stars age from 0 to 25\,Myr, they emit increasingly hard ionising radiation (Fig.~\ref{fig:bpass}) and they become situated in increasingly diffuse regions of gas. This causes the average age of stars responsible for ionising the gas to increase with decreasing $n_{\rm e}$, and therefore the hardening of ionising radiation with decreasing $n_{\rm e}$ (Fig.~\ref{fig:radiation_hardness_overall}).

\item The line ratio trends in the DIG can be explained as a direct consequence of ongoing star-formation, rather than needing to invoke a secondary ionisation source such as HOLMES. Variation in the precise star-formation history over the last 25\,Myr may explain the scatter in the observed line ratio trends between galaxies.

\item Once the \HII region around a young star has been cleared, it is able to ionise gas on galaxy-wide scales (Fig.~\ref{fig:age_dist_freq}). This provides a plausible explanation for the relative uniformity (relatively small scatter) of line ratio trends radially within individual galaxies, after accounting for metallicity gradients.
\end{enumerate}

We have shown how careful modelling of emission line properties of simulated galaxies can be compared to high-resolution extragalactic observations in order to constrain stellar feedback models and stellar SEDs. Both the observational and modelling capabilities have only recently become sufficiently advanced for this task, opening an exciting avenue to probe the intricate interactions between stars and the ISM in detail.

\section*{Acknowledgements}

We thank Debora Sijacki and Martin Haehnelt for valuable feedback. WM thanks the Science and Technology Facilities Council (STFC) Center for Doctoral Training (CDT) in Data intensive Science at the University of Cambridge (STFC grant number 2742968) for a PhD studentship. RM and WM acknowledge support by the Science and Technology Facilities Council (STFC), by the ERC through Advanced Grant 695671 ``QUENCH'', and by the UKRI Frontier Research grant RISEandFALL. RM also acknowledges funding from a research professorship from the Royal Society. FB acknowledges funding from the INAF — Fundamental Astrophysics 2022 research program. HL acknowledges fundings from the National Key R\&D program of China No. 2023YFB3002502 and the National Natural Science Foundation of China under No. 12373006.

\section*{Data Availability}

Raw data were generated by performing simulations at the NASA Pleiades computer. Derived data supporting the findings of this study are available from the corresponding author WM on request.



\bibliographystyle{mnras}
\bibliography{dig_mcclymont} 








\bsp	
\label{lastpage}
\end{document}